\documentstyle[12pt,epsf,epsfig]{article}

\title{Dynamics of One- and Two-dimensional Kinks in Bistable 
Reaction-Diffusion Equations with Quasi-Discrete Sources of Reaction}
\date{\today}

\author{\\
{\bf Horacio G. Rotstein}
\thanks{E-mail: horacio@cs.brandeis.edu},
{\bf Anatol M. Zhabotinsky},
{\bf Irving R. Epstein}
\\ 
Department of Chemistry and Volen Center for Complex Systems, \\
Brandeis University, MS 015, Waltham, MA 02454-9110, USA \\
}

\def\tet{\theta}
\def\bet{\beta}
\def\epsl{\epsilon}
\def\eps2{\epsilon^{2}}
\def\ep4s{\epsilon^{4}}

\def\ur{\rho_{1}}
\def\zer{\rho_{0}}

\def\inveps2{\frac{1}{\epsilon^{2}}}
\def\inv3eps{\frac{1}{\epsilon^{3}}}
\def\bnu{\bar{h}}

\def\sbet{\bet(x,s)}
\def\tbet{\bet(\rho,\tet)}
\def\xsbet{\bet_{x}(x,s)}
\def\ysbet{\bet_{y}(x,s)}
\def\rtbet{\bet_{r}(\rho,\tet)}
\def\ttbet{\bet_{\tet}(\rho,\tet)}
\def\msbet{\sbet^{\frac{1}{2}}}
\def\tsbet{\sbet^{\frac{3}{2}}}
\def\mtbet{\tbet^{\frac{1}{2}}}

\def\phit{\phi_{t}}

\def\Phit{\Phi_{t}}
\def\Phiz{\Phi_{z}}

\def\Phixx{\Phi_{xx}}
\def\Phizz{\Phi_{zz}}

\def\Phizx{\Phi_{zx}}

\def\zPhi{\Phi^{0}}
   
\def\zPhiz{\Phi_{z}^{0}}

\def\zPhizz{\Phi_{zz}^{0}}

\def\zPhizx{\Phi_{zx}^{0}}

\def\zPhixi{\Phi_{\xi}^{0}}
\def\zPhixixi{\Phi_{\xi\xi}^{0}}

\def\oPhi{\Phi^{1}}

\def\oPhizz{\Phi_{zz}^{1}}

\def\oPhixixi{\Phi_{\xi\xi}^{1}}
\def\est{s_{t}}
\def\st2{s_{t}^{2}}
\def\esx{s_{x}}
\def\sx2{s_{x}^{2}}

\def\sxx{s_{xx}}

\def\zst2{S_{0,t}^{2}}

\def\zsx2{S_{0,x}^{2}}

\def\ost2{S_{1,t}^{2}}

\def\osx2{S_{1,x}^{2}}

\def\hrt{\rho_{t}}
\def\rht2{\rho_{t}^{2}}
\def\hrc{\rho_{\theta}}
\def\rhc2{\rho_{\theta}^{2}}
 
\def\rhcc{\rho_{\theta\theta}}

\def\zrht{\rho_{t}^{0}} 
\def\zrht2{\rho_{t}^{2}^{0}}
\def\zrhc{\rho_{\theta}^{0}}
\def\zrhc2{\rho_{\theta}^{2}^{0}}

\def\orht{\rho_{t}^{1}}
\def\orht2{\rho_{t}^{2}^{1}}
\def\orhx{\rho_{\theta}^{1}}
\def\orhx2{\rho_{\theta}^{2}^{1}}

\def\ar2{R_{0}^{2}}

\def\art2{R_{0,t}^{2}}

\def\lap-phi{\Delta \phi}
\def\lapt-phi{\Delta \phit}

\def\xic{1 + \sx2}
\def\axic{(\xic)^{\frac{1}{2}}}
\def\bxic{(\xic)^{\frac{3}{2}}}
\def\xip{\rho^{2} + \rhc2}

\def\t*{t_{*}}
\def\begeq{\begin{equation}}
\def\endeq{\end{equation}}
\def\begdis{\begin{displaymath}}
\def\enddis{\end{displaymath}}
\def\bk{\bigskip}

\def\nd{\noindent}

\begin{document}

\maketitle

\begin{abstract}

	We study the evolution of fronts in a bistable reaction-diffusion
system when the nonlinear reaction term is spatially
non-homogeneous. This equation has been used to model wave propagation
in various biological systems. Extending previous works on homogeneous 
reaction terms, we derive asymptotically an equation governing the
front motion, which is strongly nonlinear and, for the 
two-dimensional case, generalizes the classical mean curvature flow
equation. We study the motion of one- and two- dimensional fronts, finding that
the non-homogeneity acts as a "potential function" for the motion of
the front; i.e., there is wave propagation failure and the steady
state solution depends on the structure of the function describing the
non-homogeneity.  

\end{abstract} 

\nocite{kn:flakla1}
\nocite{kn:flawil1}
\nocite{kn:flomaz1}
\nocite{kn:fuklee1}
\nocite{kn:leeric1}
\nocite{kn:cop1}
\nocite{kn:bishor1}
\nocite{kn:mitkla1}
\nocite{kn:mitkla2}
\nocite{kn:mit1}
\nocite{kn:klamit1}
\nocite{kn:ponkei1}
\nocite{kn:keismi1}
\nocite{kn:peapon1}
\nocite{kn:gru1}
\nocite{kn:gru2}
\nocite{kn:kee1}
\nocite{kn:kee2}

\section{Introduction}

\nd In this paper we consider the following equation

\begeq
	\phit = D\ \lap-phi + \alpha\ \bet(x,y)\ [ f(\phi) + h ],
							\label{eq:def1}
\endeq

\nd in a bounded region, \( \Omega \subset R^{n} \), \( n = 1, 2 \), with 
smooth boundary \( \partial \Omega \) for Neumann boundary conditions on 
\( \partial \Omega \). The positive constants \( D \) and \( \alpha \)
are the diffusion coefficient and the production rate of the reactants,
respectively. The function \( f \) is a bistable function 
(the derivative of a double well potential); i.e., a real odd function with 
positive maximum equal to \( \phi^{\ast} \), negative minimum equal to 
\( -\phi^{\ast} \) and precisely three zeros in the closed interval 
\( [a_{-},a_{+}] \) located at \( a_{-} \), \( a_{0} \) and 
\( a_{+} \). For simplicity and without lost of generality we will consider 
in our analysis \( a_{-} = -1 \), \( a_{0} = 0 \) and \( a_{+} = 1 \). The 
prototype example is \( f(\phi) = (\phi - \phi^{3}) / 2 \). 
The constant \( h \) in (\ref{eq:def1}), 
assumed to be small in absolute value, specifies the difference of the 
potential minima of the system as will be explained later. Altough the 
analysis presented below 
will be valid for a general class of positive differentiable function 
\( \beta \), we have in mind some particular cases which are described below. 
In what follows \( \eta \) is a positive constant.
\bk
 
\nd Case 1) There is a sequence of points on the real line, \( x_{k} \),
\( k = 1, \ldots, N \), with \( N \) finite or infinite, where the function 
\( \beta \) reaches a maximum,
 
\begeq
	\bet(x) = \sum_{k=1}^{N}\ e^{-\eta (x-x_{k})^{2}}.
							\label{eq:def1b1}
\endeq

\nd Case 2) There is a sequence of lines in the plane, \( y_{k} \), 
\( k = 1, \ldots, N \), with \( N \) finite or infinite, where the function 
\( \beta \), independent of \( x \), reaches a maximum,

\begeq
	\bet(x,y) = \sum_{k=1}^{N}\ e^{-\eta (y-y_{k})^{2}}.
							\label{eq:def1b2}
\endeq

\nd Case 3) There is a sequence of points in the plane, \( (x_{k},y_{j}) \), 
\( k = 1, \ldots, N \), \( j = 1, \ldots, M \) with \( N \) and \( M \) finite 
or infinite, where the function \( \beta \) reaches a maximum,

\begeq
	 \bet(x,y) = \sum_{k=1}^{N} \sum_{j=1}^{M} 
	\sigma(x-x_{k},y-y_{j};\eta), 
	\ \ \ \ \ \ \ \ \ \ {\mbox where}\ \ \ \ \ \ \ \ \ \
	\sigma(x,y;\eta) = e^{-\eta(x^{2}+y^{2})}.
							\label{eq:def1c}
\endeq

\nd Case 4) There is a sequence of circles in the plane, \( \rho = \rho_{k} \),
\( k = 1, \ldots, N \), with \( N \) finite or infinite, and where \( \rho \) 
represents the radial polar coordinate, where the function 
\( \beta \) reaches a maximum,

\begeq
	\bet(\rho) = \bet(x,y) = \sum_{k=1}^{N}\ 
	e^{-\eta (\rho-\rho_{k})^{2}}.
							\label{eq:def1g}
\endeq
\bk

\nd For sufficiently large values of \( \eta \), \(\beta \) as 
given by (\ref{eq:def1b1}-\ref{eq:def1g}) are approximations of distributions
of discrete sources of reaction. We will refer to the points \( x_{k} \) and  
\( (x_{k},y_{j}) \),\( k = 1, \ldots, N \), \( j = 1, \ldots, M \) as   
quasi-discrete sources of reaction or quasi-discrete (QD) sites
and to the stripes \( y = y_{k} \) and circles \( \rho = \rho_{k} \) 
\( k = 1, \ldots, N \) as quasi-semi-discretes sources of reaction or 
quasi-semi-discrete (QS) sites. We define \( d \) to be the shortest distance 
between two adjacent QD or QS sites. 
\bk

\nd In the last several years, partial differential equations with nonlinear 
discrete sources of reaction (NDSR) have been used to model  
phenomena in different fields ranging from physics to biology,  
including the study of pinning in the dislocation motion in 
crystals, breathers in nonlinear crystal lattices, Josephson junction arrays 
and the biophysical description of calcium release waves 
\cite{kn:flakla1}-\cite{kn:kee2}. The non-homogeneous  version of the 
bistable equation with discrete sources of reaction

\begeq
	\phit = D\ \lap-phi + \alpha\ \sum_{k} \delta(x-x_{k})\ 
	[ f(\phi) + h ],
							\label{eq:intro1}
\endeq

\nd has received special attention for its applicability to the dynamics 
of charge density waves \cite{kn:fuklee1},
\cite{kn:cop1}-\cite{kn:mitkla1}, \cite{kn:gru1,kn:gru2} and to the
dynamics of calcium release waves
\cite{kn:mitkla1,kn:kee1,kn:kee2}. Equation 
(\ref{eq:intro1}) describes the evolution of some 
concentration (in chemical or biological applications) or order parameter 
(in some physical applications) \( \phi \) in a discrete array of nonlinear 
reaction sites embedded in a continuum. In \cite{kn:kee2} equation 
(\ref{eq:def1}) with an additional term on the right side, \( - a\ \phi \),
and \( \beta \) given by (\ref{eq:def1b1}) has been used to model calcium 
release and uptake in cardiac cells via ryanodine receptors. In this model
\( \phi \) represents the concentration of \( Ca^{2+} \) and \( f(\phi) \) 
represents the calcium-induced calcium release (CICR) activity of the release 
mechanism. When \( a = 1 \), the model allows for continuous spatial uptake, 
whereas when \( a = 0 \), it is assumed that release and uptake both occur at 
the QD or QS sites. In \cite{kn:mitkla2} the function \( f(\phi) \) was taken 
to be the derivative of a sine-Gordon potential; i.e., 
\( f(\phi) = - \sin(\phi) \). Note that this last function is equal to the 
derivative of a double well potential, as described above, in a restricted 
domain of definition. 
\bk

\nd When the function \( \sum_{k} \delta(x - x_{k}) \) in (\ref{eq:intro1}) is
replaced by a constant, say \( 1 \), then by appropriate rescaling we have 
the bistable equation

\begeq
	\phit = b\ \lap-phi + f(\phi) + h,
							\label{eq:def1a}
\endeq

\nd which describes a phase transition dynamics process,where \( \phi
\) is a non-conserved order parameter. Note that for the particular cases 
\( f(\phi) = (\phi - \phi^{3})\ /\ 2 \) and \( f(\phi) = sin(\phi) \)
equation (\ref{eq:def1a}) is the Ginzburg-Landau equation and the
overdamped sine-Gordon equation respectively. Equation
(\ref{eq:def1a}) can be derived by considering a
physical system 
whose free energy 
is assumed to be of the form

\begeq
	F_{b}(\phi) = \int_{\Omega} \left( \frac{b}{2} 
	( \nabla \phi )^{2} + F(\phi) - h\, \phi \right) dx,
							\label{eq:def2}
\endeq   

\nd where \( F \) is a double well potential having the two equal minima.
Note that \( F(\phi) - h\ \phi \) is a double-well potential 
with one local minimum and one global minimum. The functional derivative of 
(\ref{eq:def2}) is given by

\begeq
	\frac{\delta F_{b}}{\delta \phi} = - b\ \Delta \phi -
	f(\phi) - h,
							\label{eq:def3}
\endeq

\nd where \( f(\phi) = - F'(\phi) \). The right side of equation 
(\ref{eq:def3}) may be considered as a generalized force indicative of the 
tendency of the free energy to decay towards a minimum. The bistable equation
(\ref{eq:def1a}) is obtained by assuming that \( \phi \) decreases at
a rate proportional to that generalized force.
\bk 

\nd Equation (\ref{eq:def1a}) for \( n = 1 \) possesses a travelling kink 
solution moving with velocity proportional to \( h \). A kink is a solution 
that connects the two local minima of the double well potential. If 
\( h \neq 0 \) then the kink propagates from the locally stable minimum to the
globally stable minimum. Of special interest are kinks in which the transition 
between the two minima takes place in a region of order of magnitude 
\( \epsl \ll 1 \); i.e., the kinks have rapid spatial variation between the 
two ground states. For this case, the point on the line (for \( n = 1 \))
or the set of points in the plane (for \( n = 2 \)) for which the order 
parameter \( \phi \) vanishes are called the interface or the front.
Allen and Cahn \cite{kn:allcah1} and Rubinstein, Sternberg and Keller 
\cite{kn:rubste1} showed that for (\ref{eq:def1a}) with \( b \ll 1 \)
and \( h = 0 \) curved fronts in the plane move with normal velocity 
proportional to their curvature, according to the FMC (flow by mean
curvature) equation 

\begeq
	\frac{\est}{\axic} = \frac{\sxx}{\bxic} - \bnu,		
							\label{eq:flwmc}
\endeq

\nd where \( y = s(x,t) \) is the Cartesian description of the interface in 
the plane, and \( \bnu \) is proportional to \( h \) (see also \cite{kn:fif1}).
For a circular interface and \( \bnu = 0 \) (both phases have equal potential),
the curvature is the reciprocal of the 
radius \( R \), and (\ref{eq:flwmc}) becomes \( R_{t} = - \frac{1}{R(t)} \),
whose solution satisfying \( R(0) = R_{0} \) is given by
\( R(t) = \sqrt{ R_{0}^{2} - 2 t } \); i.e., circles shrink to a  
point at a critical time 
\( t_{c,0} = \frac{R_{0}^2}{2}\). For \( \bnu > 0 \) the value of the
critical time decreases; i.e., \( t_{c,\bnu} < t_{c,0} \). For \( \bnu
< 0 \) there exists a critical value \( \bnu_{c} \) such that if 
\( \bnu > \bnu_{c} \), then circles still shrink to a point in a finite
time \( t_{c,\bnu} < t_{c,0} \), whereas if \( \bnu < \bnu_{c} \), then
circles grow unboundedly.
For more general shapes, such an expression for the distance of every point in 
the interface from the origin is difficult to obtain, but there are some 
analytical results showing that the behaviour is similar.
Gage and Hamilton \cite{kn:gagham1} proved that (\ref{eq:flwmc}) 
shrinks convex curves embedded in the plane \( R^{2} \) to a point. They
showed that such curves remain convex and become asymptotically circular as 
they shrink. Grayson \cite{kn:gra1} extended this result to a more general case
showing that embedded curves become convex without developing singularities; 
i.e., curve shortening shrinks embedded plane curves smoothly to points, with 
round limiting shape.
\bk

\nd On rescaling (\ref{eq:intro1}) \( x \rightarrow x / d \) and 
\( t \rightarrow \alpha\ t \) and defining \( b = D / \alpha\ d^{2} \)
(for n = 1 ) \cite{kn:mit1}, equation (\ref{eq:intro1}) becomes

\begeq
	\phit = b\ \lap-phi + \sum_{k} \delta(x-x_{k})\ [f(\phi) + h].
							\label{eq:intro2}
\endeq

\nd The parameter \( b \) can be thought of as a measure of how close  
(\ref{eq:intro2}) is to its continuous limit (\ref{eq:def1a}). If 
\( b \rightarrow \infty \) equation (\ref{eq:intro2}) behaves like its 
continuous counterpart (with the corresponding rescaling); i.e., it possesses 
a travelling kink solution moving with velocity proportional to \( h \) 
\cite{kn:mit1}. For \( b \) small, Mitkov et al. \cite{kn:mitkla1,kn:mit1} 
have found numerically that the front dynamics results in burst waves 
characterized by time periodicity in a frame moving along with the front. 
For \( b \) small enough, the wave no longer propagates, but relaxes to a 
stationary kink; i.e., the waves are pinned. 
\bk

\nd For the one-dimensional version of (\ref{eq:def1}), with \( \beta \) 
given by (\ref{eq:def1b1}), as well as for the continuous spatial uptake 
version, \( a = 1 \), Keener \cite{kn:kee2} demonstrated the failure of 
wavefront propagation if the separation between QD sites is large enough.
\bk

\nd For (\ref{eq:def1}) we define the spatial and temporal
dimensionless variables as in \cite{kn:mit1} 

\begeq
	\hat{x} = \frac{x}{d},\ \ \ \ \ \ \ \ \ \
	\hat{y} = \frac{y}{d},\ \ \ \ \ \ \ \ \ \
	\hat{t} = \alpha\ t,
							\label{eq:def1d}
\endeq

\nd and	we also define the following dimensionless parameters

\begeq
	\epsilon = \frac{1}{d}\ \sqrt{\frac{D}{\alpha}}, \ \ 
	\ \ \ \ \ \ \ \ \ \
	\hat{\eta} = \eta\ d^{2},\ \ 
	\ \ \ \ \ \ \ \ \ \
	\hat{h} = \frac{h}{\epsilon}.
							\label{eq:def1e}
\endeq

\nd Substituting (\ref{eq:def1d}) and (\ref{eq:def1e}) into
(\ref{eq:def1}), dropping the \( \hat{} \) from the variables and parameters 
and further rescaling the time variable we obtain

\begeq
	\eps2\ \phit = \eps2\ \lap-phi + \bet(x,y)\ [ f(\phi) + \epsl\
	h ],
							\label{eq:def1f}
\endeq
 
\nd We will consider the case \( 0 < \epsl \ll 1 \); i.e., when
diffusion is slow, \( d \) is large or reaction is fast.
\bk

\nd In Section 2 we make a formal asymptotic analysis to derive an equation of
motion for the front in equation (\ref{eq:def1f}), which in Cartesian 
coordinates reads

\begeq
	\est = \frac{\sxx}{\xic} +  \frac{\esx\ \xsbet}{2\ \sbet} -
	\frac{\ysbet}{2\ \sbet} - \msbet\ \bnu
							\label{eq:flwmcextc}
\endeq

\nd where the parameter \( \bnu \), which is proportional to \( h \), is
defined later. Note that equation (\ref{eq:flwmcextc}) expressed in
polar coordinates reads

\begeq
	\hrt = 	\frac{\rhcc\ \rho - 2\ \rhc2 - \rho^{2}}{\rho\ (\xip)} + 
	\frac{\hrc}{\rho^{2}}\ \frac{\ttbet}{2\ \tbet} -
        \frac{\rtbet}{2\ \tbet} - \mtbet\ \bnu.
							\label{eq:flwmcextp}
\endeq

\nd This equation generalizes the FMC equation (\ref{eq:flwmc}) with a strong 
nonlinearity accounting for the influence of the function \( \beta \) on the 
front motion. The method we use is the same as that used in 
\cite{kn:rotnep1} for the study of the evolution of kinks in the nonlinear wave
equation. We present it here in some detail for the sake of completeness. 
In Section 3 we
study the evolution of one-dimensional fronts by means of (\ref{eq:flwmcextc}).
We show that for \( \bnu = 0 \) the function \( \beta \) acts as a "potential
function" for the motion of the front; i.e., a front initially placed 
between two maxima of \( \beta \) asymptotically approaches the intervening
minimum, unlike the classical homogeneous equation 
(\ref{eq:def1a}), for which fronts, whose motion is governed by 
(\ref{eq:flwmc}), move with a velocity proportional to \( \bnu \). This 
result is a consequence of the non-homogeneity of the nonlinear reaction 
term. In Section 4 we study the evolution of two-dimensional fronts by means
of (\ref{eq:flwmcextc}). We show analytically that a radially symmetric and 
non-constant function \( \beta \) stabilizes a circular domain of one phase 
inside the other phase analogous to the one-dimensional case. 
This behavior, also arised as a consequence of the 
non-homogeneity of the nonlinear reaction term. The evolution of closed 
curves according to (\ref{eq:flwmcextc}) for \( \beta \) given by
(\ref{eq:def1c}) is studied numerically. We observe that closed convex curves
evolve to a final shape determined by \( \beta \). Our conclusions appear in  
Section 5. 

\section{Asymptotic Analysis: Derivation of the Equation of Front Motion}

\nd We assume that for small \( \epsilon \geq 0 \) and all 
\( t \in [0,T] \), the domain \( \Omega \) can be divided into two  
open regions \( \Omega_{+}(t;\epsilon) \) and \( \Omega_{-}(t,\epsilon) \) 
by a curve \( \Gamma(t;\epsilon) \), which does not intersect 
\( \partial \Omega \). This interface, defined by

\begeq
	\Gamma(t;\epsilon) := \left\{ x \in \Omega : \phi(x,t;\epsilon) = 0 
	\right\}, 
							\label{definterf}
\endeq

\noindent is assumed to be smooth, which implies that its curvature and its 
velocity are bounded independently of \( \epsilon \). We also assume that there
exists a solution \( \phi(x,t;\epsl) \) of (\ref{eq:def1}), defined for 
small \( \epsl \), for all \( x \in \Omega \) and for all \( t \in [0,T] \) 
with an internal layer. As \( \epsl \rightarrow 0 \) this solution 
is assumed to vary continuously through the interface, taken the value
\( 1 \) when \( x \in \Omega_{+}(t;\epsilon) \), \( -1 \) when 
\( x \in \Omega_{-}(t,\epsilon) \), and varying rapidly but smoothly
through the interface. By carrying out a singular perturbation 
analysis for \( \epsl \ll 1 \), we obtain the law of motion of the interface, 
treating it as a moving internal layer of width \( O(\epsilon) \). We focus on 
the dynamics of the fully developed layer, and not on the process by which it 
was generated.
\bk

\nd In Cartesian coordinates the interface is represented by 
\( y = s(x,t,\epsl) \) for \( \epsl \) sufficiently small. We assume that the
curvature of of the front is small compared to its width and define, in a 
neighborhood of the interface, a new variable

\begdis
	z := \frac{y - s(x,t,\epsl)}{\epsl}	
\enddis

\nd which is \( {\cal O}(1) \) as \( \epsl \rightarrow 0 \). We call
\( \Phi \) the asymptotic form of \( \phi \) as \( \epsl \rightarrow 0 \) with
\( z \) fixed; i.e.,  

\begeq	
	\phi = \Phi(z,x,t,\epsl).			\label{eq:as1}
\endeq

\nd The field equation (\ref{eq:def1f}) in \( (z,x,t) \) 
coordinates becomes

\begdis
	\eps2\ \Phit - \epsl\ \est\ \Phiz = \eps2\ \Phixx -
	 2\ \epsl\ \esx\ \Phizx + (1 + \sx2) \Phizz -
\enddis

\begeq
        - \epsl\ \sxx\ \Phiz +  
	\bet(x,s+\epsl\,z)\ f(\phi) + \epsl\ \bet(x,s+\epsl\,z)\ h.
							\label{eq:as2}
\endeq

\nd The asymptotic expansions of \( \Phi \) and \( S \) are assumed to 
have the form

\begdis
	\Phi \sim \zPhi + \epsl\ \oPhi + {\cal O}(\eps2), 
	\ \ \ \ \ {\mbox as}\ \ \epsl 
	\rightarrow 0.
\enddis

\nd Thus

\begdis
	\bet(x,s+\epsl\,z) = \sbet + \epsl\ z\ \ysbet + {\cal O}(\eps2).
\enddis	

\nd Substituting into (\ref{eq:as2}) and equating coefficients of 
the corresponding powers of \( \epsl \) we obtain the following problems
for \( {\cal O} (1)\) and \( {\cal O}(\epsl)\) respectively:
\bk

\begeq
	(\ \xic\ )\ \zPhizz + \sbet\ f(\zPhi) = 0,	\label{eq:prob0}
\endeq
\bk

\begdis
	(\ \xic\ )\ \oPhizz + \sbet\ f'(\zPhi)\ \oPhi = 
	( \sxx - \est ) \zPhiz + 			
\enddis

\begeq
	+ 2\ \esx\ \zPhizx - \ysbet\ z\ f(\zPhi) - \sbet\ h. 
							\label{eq:prob1}
\endeq
\bk

\nd In order to solve (\ref{eq:prob0}) we define a new variable

\begeq
	\xi := \frac{\msbet}{\axic}\ z.			\label{eq:defxi}
\endeq

\nd In terms of \( \xi \), equation (\ref{eq:prob0}) reads

\begeq
	\zPhixixi + f(\zPhi) = 0,			\label{eq:probxi0}
\endeq

\nd whose solution is \( \zPhi = \Psi(\xi) \), the unique solution of
\(\ \ \Psi'' + f(\Psi) = 0,\ \Psi(\pm \infty) = \pm 1,\ \Psi(0) = 0 \). 
Thus

\begeq
	\zPhi = \zPhi\left(\frac{\msbet}{\axic}\ z \right).   \label{eq:solxi0}
\endeq

\nd In terms of \( \xi \), \( x \) and \( t \), equation 
(\ref{eq:prob1}) reads

\begdis
	\oPhixixi + f'(\zPhi)\ \oPhi = 
	\frac{\sxx - \est}{\msbet\ \axic}\ \zPhiz + 
\enddis

\begdis
	+ 2\ \esx\ \left[ \frac{\xsbet + \ysbet\ \esx}{2\ \tsbet\ \axic} -
	\frac{\esx\ \sxx}{\msbet\ \bxic} \right]\ (\xi\ \zPhixixi + \zPhixi) 
\enddis

\begeq
	- \frac{\ysbet\ \axic}{\tsbet}\ \xi\ f(\zPhi) - h.
							\label{eq:probxi1}
\endeq

\nd It is straightforward to check that \( \Psi'(\xi) \) satisfies 
the homogeneous equation 

\begeq
	 \oPhixixi + f'(\zPhi)\ \oPhi = 0.		\label{eq:homog}
\endeq

\nd That means that the operator \( \Lambda \) defined as follows

\begeq
	\Lambda := \frac{\partial^{2}}{\partial \xi^{2}} + f'(\zPhi)
							\label{eq:operat}
\endeq

\nd has a simple eigenvalue at the origin with \( \Psi' \) as the 
corresponding eigenfunction. Then the solvability condition for the equation 
(\ref{eq:probxi1}) gives

\begdis
	\frac{\sxx - \est}{\msbet\ \axic}\ \int_{-\infty}^{\infty}
	(\Psi')^{2}\ d\xi + 
\enddis

\begdis
	2\ \esx\ \left[ \frac{\xsbet + \ysbet\ \esx}{2\ \tsbet\ \axic} -
	\frac{\esx\ \sxx}{\msbet\ \bxic} \right]\  
	\int_{-\infty}^{\infty} (\xi\ \Psi'' + \Psi')\ \Psi'\ d\xi
\enddis

\begeq
	- \frac{\ysbet\ \axic}{\tsbet}\ \int_{-\infty}^{\infty}
	\xi\ f(\Psi)\ \Psi'\ d\xi - h \int_{-\infty}^{\infty} \Psi'\
	d\xi = 0.
							\label{eq:fredholm}
\endeq

\nd A simple calculation shows that 

\begeq
	\int_{-\infty}^{\infty} \xi \Psi' \Psi'' d\xi = - \frac{1}{2} 
	\int_{-\infty}^{\infty} (\Psi')^{2} d\xi \ \ \ \ \ \ \ \ \ \
							\label{eq:integ}
	\mbox {and}\ \ \ \ \ \ \ \ \ \
	\int_{-\infty}^{\infty} \xi\ f(\Psi) \Psi'\ d\xi = \frac{1}{2} 
	\int_{-\infty}^{\infty} (\Psi')^{2} d\xi \ \ \ \ \ \ \ \ \ \
\endeq

\nd We define

\begeq
	\bnu := h\ \frac{\Psi(+\infty) - \Psi(-\infty)}{
	\int_{-\infty}^{\infty} (\Psi')^{2} d\xi},
							\label{eq:defpsi}
\endeq

\nd Substituting (\ref{eq:integ}) and (\ref{eq:defpsi}) into
(\ref{eq:fredholm}) and rearranging terms we get (\ref{eq:flwmcextc}). 
Note that for \( f(\phi) = \frac{\phi - \phi^{3}}{2} \) 
(Ginzburg-Landau theory), \( \Psi(\xi) = \tanh \frac{\xi}{2} \) and
\( \bnu = 3\ h \) whereas for \( f(\phi) = \sin \phi \) 
(overdamped sine-Gordon), \( \Psi(\xi) = 4\ \mbox{tan}^{-1} e^{\xi} - \pi \)
and \( \bnu = \frac{\pi}{4}\ h \).

\section{Front Motion in 1D}

\nd For a one-dimensional system, equation (\ref{eq:flwmcextc}) reads

\begeq
	\est = - \frac{\bet'(s)}{2\ \bet(s)} - \bet^{\frac{1}{2}}(s)\
	\bar{h}.
							\label{eq:flow1d}
\endeq

\nd We will concentrate on functions \( \bet \) of the form
(\ref{eq:def1b1}), 

\begeq
	 \bet(s) = \sum_{k=1}^{N}\ e^{-\eta(s - x_{k})^{2}},
							\label{eq:bet1d}
\endeq

\nd altough the same analysis can be done for a general differentiable
function. 
\bk

\nd In order to analyze the motion of the front we need to look at
the roots of the function 

\begeq
	g(s) = - \bet'(s) - 2\ \beta^{\frac{3}{2}}(s)\ \bar{h},
\endeq

\nd which are the equilibrium points of the interface.
As an example, in Figure \ref{discsites-1} we can see
the graph of \( \bet(s) \) and \( g(s) \) respectively for 
\( \eta = 1000 \) and \( x_{1} = 2 \), \( x_{2} = 1 \), 
\( x_{3} = 0 \) and \( x_{4} = -1 \), \( x_{5} = -2 \) and various
values of \( \bnu \). For 
\( \bar{h} = 0 \), \( g(s) \) has \( 9 \) roots in the range
considered. Five of them are \( x_{k} \), \( k = 1, \ldots, 5 \); i.e., they
correspond to the maxima of \( \bet(s) \). The other four correspond
to the minima of \( \bet(s) \): \( s_{1} \), \( s_{2} \), \( s_{3} \)
and \( s_{4} \) from left to right. It is easy to see
that \( s_{k} \), \( k = 1, \ldots , 4 \) are stable whereas 
\( x_{k} \), \( k = 1, \ldots , 5 \) are unstable. Thus one dimensional
fronts initially at non-equilibrium points \( x \) move until they reach a
stable equilibrium point; i.e., a front initially at a point 
\( x \in (x_{k},x_{k+1}) \) approaches asymptotically \( s_{k} \). 
Fronts starting initially at \( x > x_{1} \) or \( x < x_{N} \) will
move forever. This behavior is in contrast with the classical FMC case 
(\ref{eq:flwmc}), where one dimensional fronts move only if 
\( \bnu \neq 0 \). In order to understand the behaviour of \( g(s) \) as 
\( \bnu \) increases above zero we can look at a function \( \beta(s)  \)
with a single peak at \( x_{1} = 0 \); i.e., \( \bet(s) = e^{-\eta s^{2}} \). 
This function will approximate every (\ref{eq:def1b1}) if \( \eta \gg 1 \), 
so that the influence of peaks on one another is very small. In this case
\( g(s) = 2 e^{-\eta\ s^{2}} [\eta\ s - \bnu\ e^{-\frac{\eta s^{2}}{2}}] \). 
For \( \bnu = 0 \), \( g(s) \) vanishes at \( \hat{x} = x_{1} = 0 \)  and it
is positive for \( x > 0 \) and negative for \( x < 0 \) (see
Fig. \ref{discsites-1}-b). As \( \bnu \) moves from zero, 
\( \hat{x} \), the root of \( g(s) \), will be given by the solution of 
\(\eta\ s - \bnu\ e^{-\frac{\eta s^{2}}{2}} = 0 \), an equation that has a
solution as long as \( \eta \) is sufficiently large and \( h = {\cal O}(1) \).
If \( \bnu > 0 \), \( \hat{x} > 0 \), \( g(s) \) is positive for 
\( x > \hat{x} \) and negative for \( x < \hat{x} \). If \( \bnu < 0 \) 
\( \hat{x} < 0 \). We can see the shape of \( g(s) \) as \( \bnu \) increases 
in Figure
\ref{discsites-1}. In summary, as \( \bnu \) increases or decreases the
behavior of the front is similar to the case \( \bnu = 0 \) in
contrast with the classical FMC case (\ref{eq:flwmc}) where fronts
move with a velocity proportional to the value of \( \bnu \).  
As an illustration, in Figure \ref{front-1}-a and \ref{front-1}-c we
can show the graphs of \( s_{t} \) as function of \( s \) and of \( s \)
as a function of \( t \), respectively, for \( \eta  = 30 \) and \( \bnu
= 0, 0.5 \) and \( 1 \). In Figure \ref{front-1}-b we can
see the corresponding graph of \( \bet \) as a function of \( s \). We
observe that the velocity of the front initially increases and
then decreases as the front ``leaves'' the area of the
peak of \( \bet(s) \). As \( h \) increases, the velocity decreases
for a given value of \( s \), as does the asymptotic value of \( s \).

\section{Front Motion in 2D}

\nd In this section we present some analytical and numerical results for the
front motion of closed curves in the plane according to (\ref{eq:flwmcextc}). 
The analysis of front motion in two dimensions according (\ref{eq:flwmcextc})
with a function \( \beta \) of type (\ref{eq:def1b2}) reduces to the 
analysis of front motion on a line and we shall not consider this case 
further.

\subsection{Radial Symmetry}

\nd For radially symmetric functions, \( \bet = \bet(\rho) \), and
initial fronts, equation (\ref{eq:flwmcextp}) reads

\begeq
	\hrt = -\frac{1}{\rho} - \frac{\bet'(\rho)}{2\ \bet(\rho)} - 
	\bet^{\frac{1}{2}}(\rho)\ \bar{h}.
							\label{eq:flowrs}
\endeq
	
\nd As in the previous Section, the analysis presented here can be performed
for a general differentiable positive function \( \bet \), though here
we concentrate on a function \( \bet \) of the type (\ref{eq:def1g}).
\bk


\nd We begin our analysis for the case \( N = 1 \); i.e., \(
\bet(\rho) = e^{-\eta(\rho - \ur)^{2}} \), and \( \bnu = 0 \). For
this case, equation (\ref{eq:flowrs}) becomes 

\begeq
	 \hrt = -\frac{1}{\rho} + \eta\ (\rho - \ur).
							\label{eq:case1a-rs}
\endeq

\nd Equation (\ref{eq:case1a-rs}) has only one equilibrium point, 
\( \hat\rho = \left( \eta\ \ur + \sqrt{\eta^{2} \ur^{2} + 4\ \eta}
\right) / 2\ \eta \), which is unstable. Note that \( \hat\rho
\rightarrow \ur \) as \( \eta \rightarrow \infty \). Thus, circles
with initial radius \( \zer < \hat\rho \) shrink to a point in finite 
time \( t_{c} \) whereas circles for which \( \zer > \hat\rho \) grow
unboundedly. This is in contrast with the classical case
(\ref{eq:flwmc});i.e., for a constant \( \bet \), where circles with
any initial radius shrink to a point in finite time unless \( \bnu \neq
0 \) (see introduction). For \( \ur = 0 \), \( \rho(t) = 
\sqrt{1 / \eta + (\zer^{2} - 1 / \eta)\ e^{2 \eta t}} \) and then
\( t_{c} = - \ln (1 - \eta \zer^{2}) / 2 n \).
\bk

\nd For \( N > 1 \) we need to look at the roots of the
function

\begeq
	f(\rho) = - 2\ \bet(\rho) - \rho\ \bet_{\rho}(\rho)  
                  - 2\ \rho\ \bet^{\frac{3}{2}}(\rho)\ \bnu. 
							\label{eq:case1b-rs}
\endeq

\nd As an illustration, consider Figure \ref{discsites-2}, which
shows the graphs of \( \bet(\rho) \) and \( f(\rho) \) for 
\( N = 5 \), \( \bnu = 0 \) and \( \eta = 100, 30, 11 \) and \( 4
\). In Figures \ref{discsites-2}-a and \ref{discsites-2}-b, we 
observe that \( f(\rho) \) vanishes \( 9 \) times; i.e., equation 
(\ref{eq:flowrs}) has \( 9 \) equilibrium points which are such
that \( f(\rho) \) vanishes near the maxima or minima of \(
\bet(\rho) \) with the exception of the first maximum. We call 
\( \rho_{M,k} \), \( k = 1, \ldots, 5 \) and \( \rho_{m,k} \), 
\( k = 1, \ldots, 4 \) the odd and even equilibrium points of
(\ref{eq:flowrs}), respectively, starting from that of lower value.
We can see that  \( \rho_{M,k} \), 
\( k = 1, \ldots, 5 \) are unstable whereas \( \rho_{m,k} \), 
\( k = 1, \ldots, 4 \) are stable. Thus circles with an initial radius
\( \rho_{M,k} < \zer < \rho_{M,k+1} \), \( k = 1, \ldots, 4 \) grow or
shrink to a circle of radius \( \rho_{m,k} \), circles with an initial
radius \( \zer < \rho_{M,1} \) shrink to a point in finite time and
circles with an initial radius \( \zer > \rho_{M,5} \) grow
unboundedly. This analysis can be generalized for any value of \(
\eta \). For a large number of sites \( \rho_{k} \) the analysis
would be similar. We conclude that the function \( \bet \) stabilizes a
circular domain of one phase inside the other in contrast with the
clasical case (\ref{eq:flwmc}) where circles of any initial radius
shrink to a point in finite time. In Figures  \ref{discsites-2}-c and 
\ref{discsites-2}-d we see that for lower values of \( \eta \)
some of the equilibrium points disappear (a saddle node bifurcation of stable
and unstable equilibrium points occurs). For \( \eta = 0 \) we expect no
equilibrium points, since this corresponds to the classical 
equation (\ref{eq:flwmc}). In Figure \ref{discsites-3} we have the 
graphs of \( \bet(\rho) \) and \( f(\rho) \) for 
\( N = 5 \), \( \eta = 0 \) and \( \bnu = 0, 2 \) and \( -2 \)
respectively. We observe
the influence of \( \bnu \) on the graph of 
\( f(\rho) \); i.e., on the equilibrium points of (\ref{eq:flowrs}) and the 
velocities of circles growing or shrinking. For sufficiently large
values of \( \bnu \) some equilibrium points will eventually dissapear.    
\bk 

\subsection{Some Numerical Results for More General Cases}  

\nd In this section we present some numerical results For the evolution of 
fronts according to either (\ref{eq:flwmcextc}) or (\ref{eq:flwmcextp}) with 
\( \bnu = 0 \). In all cases the function  \( \beta(x,y) \) is given by 
(\ref{eq:def1c}) with \( \eta = 10 \), \( N = M = 5 \) with sites 
\( (x_{k},y_{j}) \), \( k, j = -2, \ldots, 2 \) and 
\( (x_{0},y_{0}) = (0,0) \). In Figure \ref{front-2d-04} we show a graph
of this \( \beta \).
\bk

\nd In Figure \ref{front-04} we see the evolution of a circle of radius
\( 2 \) for \( t = 0, 0.15, 0.3 \). Comparing with Figure \ref{front-2d-04} 
we can see that the initial points \( A=(2,0), B=(0,2), C=(-2,0) \) and 
\( D=(0,-2) \) are relative maxima of \( \beta \), while \( E, F, G \) and 
\( H \), the points of intersection between the front and the lines 
\( y = x \) and \( y = -x \), lie very near relative minima of \( \beta \).
As the front evolves, \( A \), \( B \), \( C \) and \( D \) move towards 
points between the maxima of \( \beta \) while \( E \), \( F \), \( G \) and 
\( H \) remain nearly stationary. The front seeks a position along the 
minimum of \( \beta \). The same behaviour can be seen in 
Figure \ref{front-05} for the ellipse \( \frac{x^{2}}{4} + y^{2} = 1 \).
Because of the very different initial conditions, the final front differs 
front that of Figure \ref{front-04}. In Figure \ref{front-06}, we arrive
at the same final front as in Figure \ref{front-04}. In Figure 
\ref{front-06}-a \( A = (1,0) \), \( B = (0,2) \), \( C = (-1,0) \) and 
\( D = (0,-2) \) whereas in Figure \ref{front-06}-c  \( A = (1.5,0) \), 
\( B = (0,1.5) \), \( C = (-1.5,0) \) and \( D = (0,-1.5) \).

\section{Conclusions}

\begin{figure}[ph]
\begin{tabular}{llllllll}
\large \bf (a) \rm \normalsize			&	
\epsfig{file=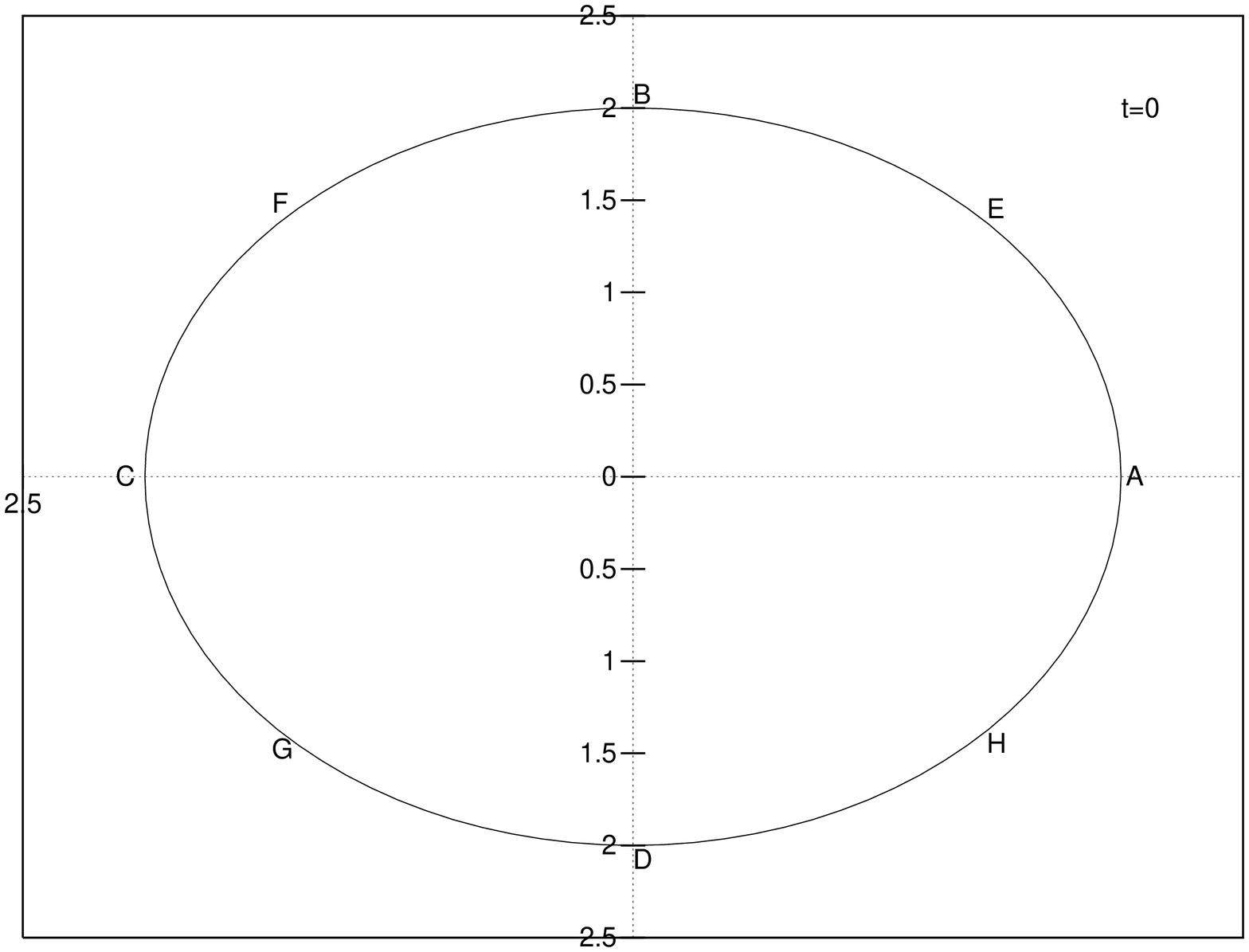,height=8cm,width=6cm,angle=-90}  \\
\large \bf (b) \rm \normalsize			&	
\epsfig{file=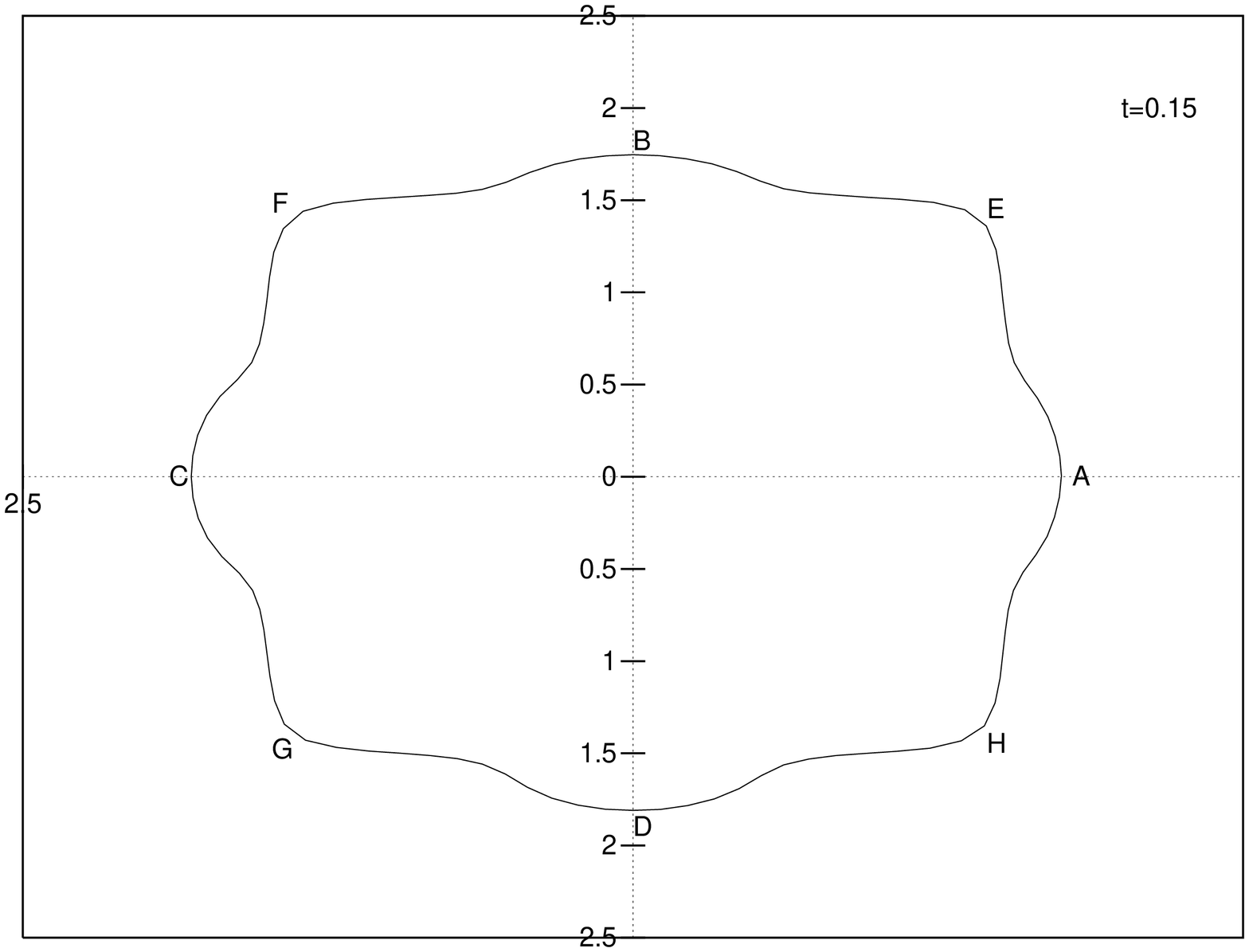,height=8cm,width=6cm,angle=-90}  \\
\large \bf (c) \rm \normalsize			&	
\epsfig{file=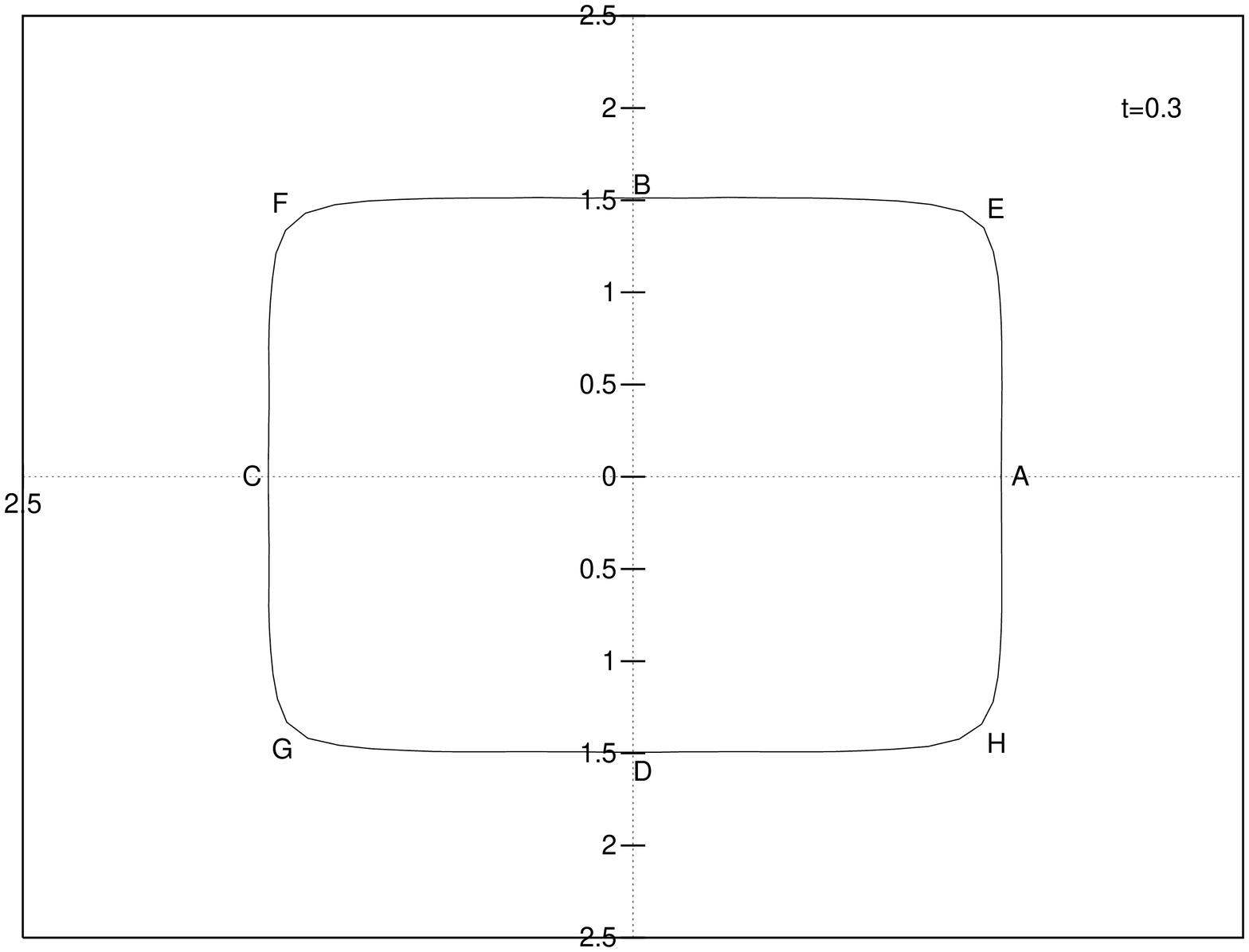,height=8cm,width=6cm,angle=-90}  
\end{tabular}
\caption{}
\label{front-04}
\end{figure}

\begin{figure}[ph]
\begin{tabular}{llllllll}
\large \bf (a) \rm \normalsize			&	
\epsfig{file=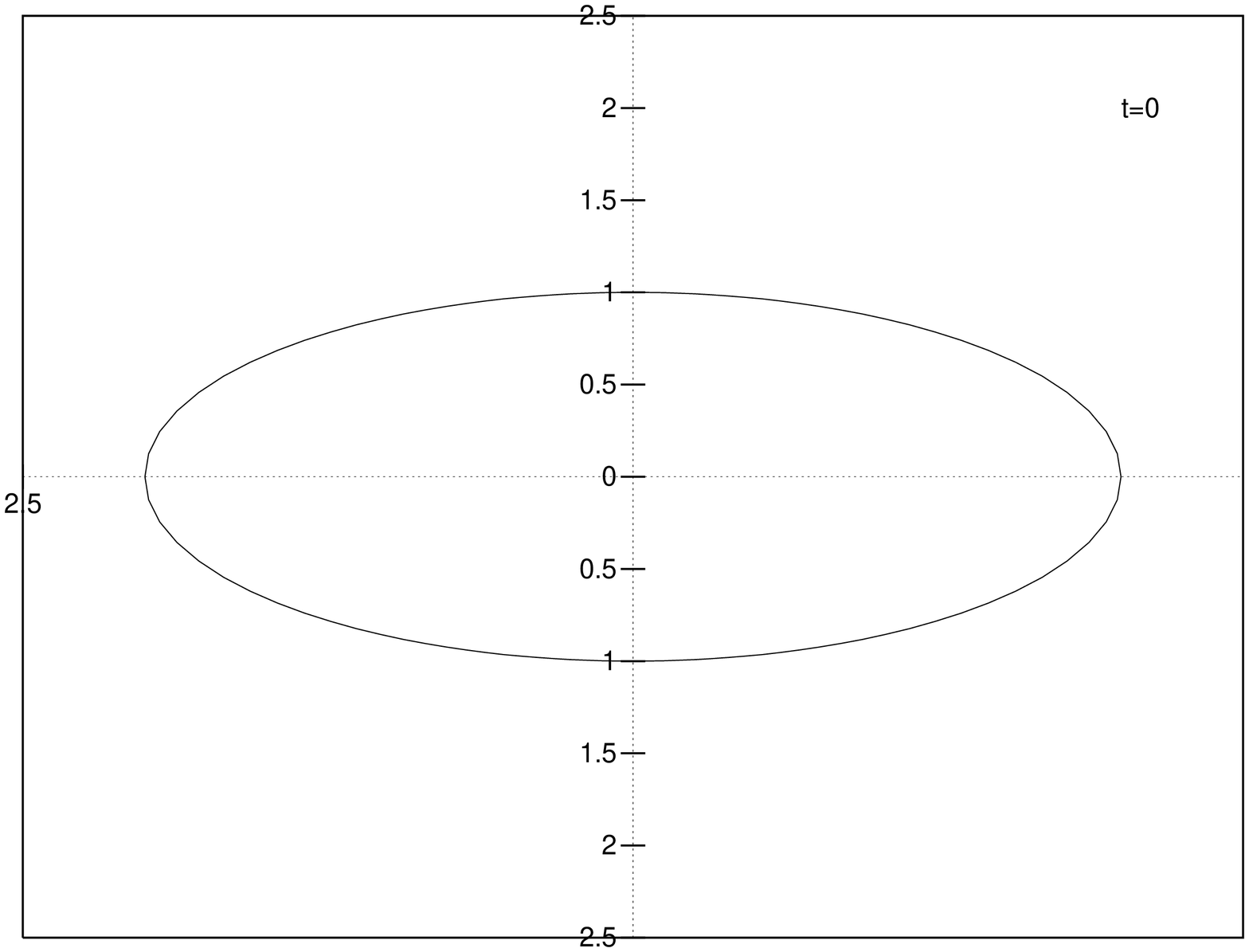,height=8cm,width=6cm,angle=-90}  \\
\large \bf (b) \rm \normalsize			&	
\epsfig{file=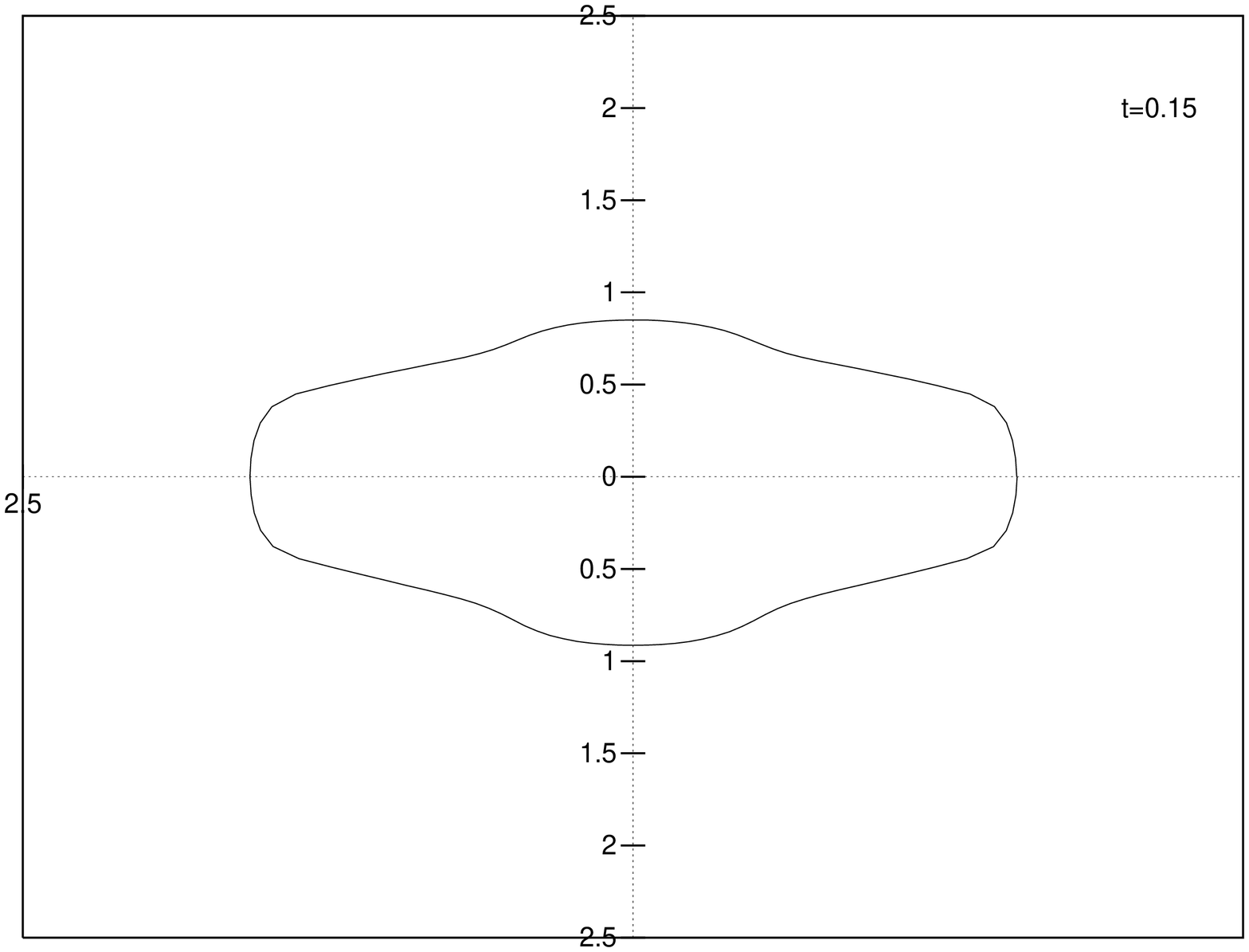,height=8cm,width=6cm,angle=-90}  \\
\large \bf (c) \rm \normalsize			&	
\epsfig{file=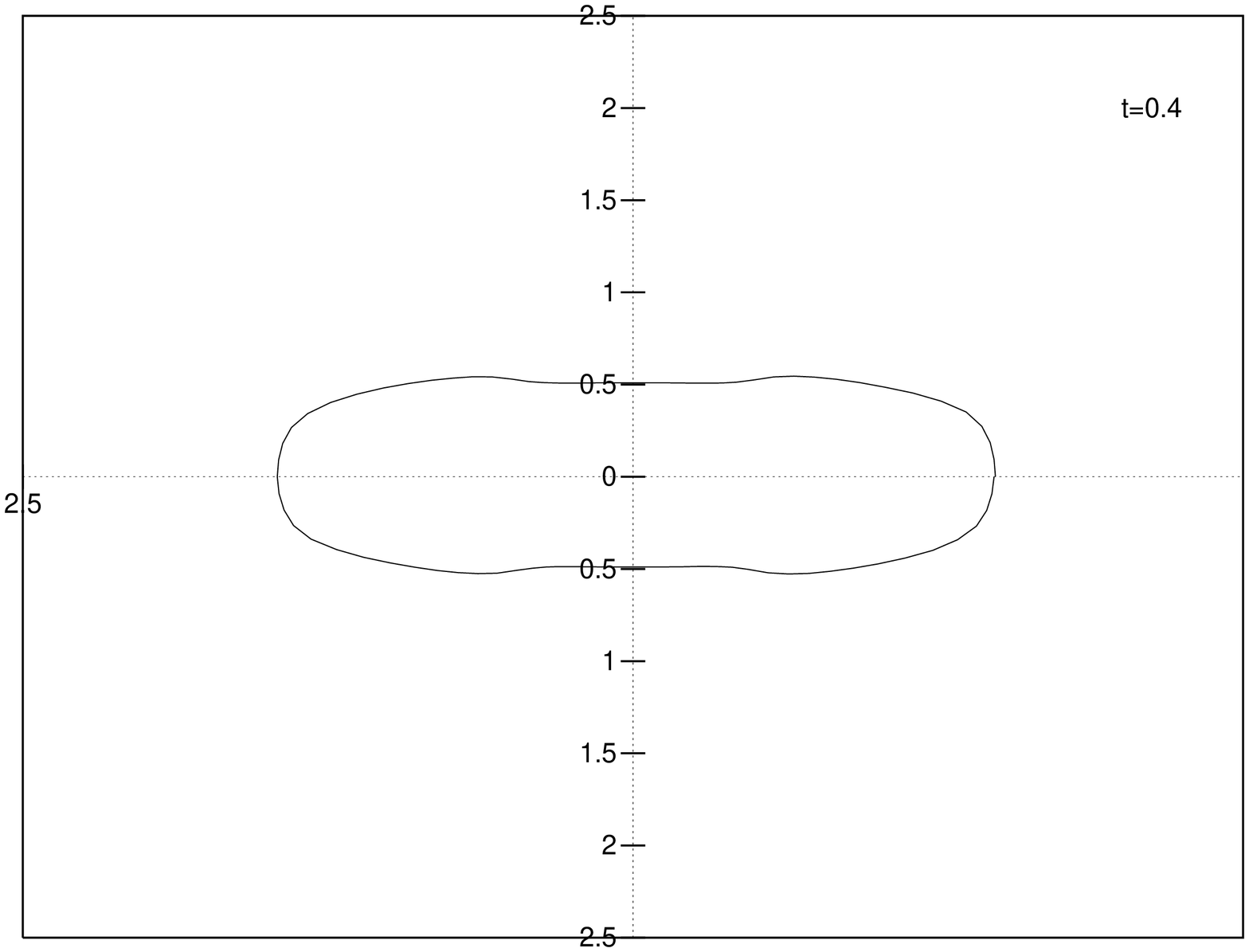,height=8cm,width=6cm,angle=-90}  
\end{tabular}
\caption{}
\label{front-05}
\end{figure}

\begin{figure}[ph]
\begin{tabular}{llllllll}
\large \bf (a) \rm \normalsize			&	
\epsfig{file=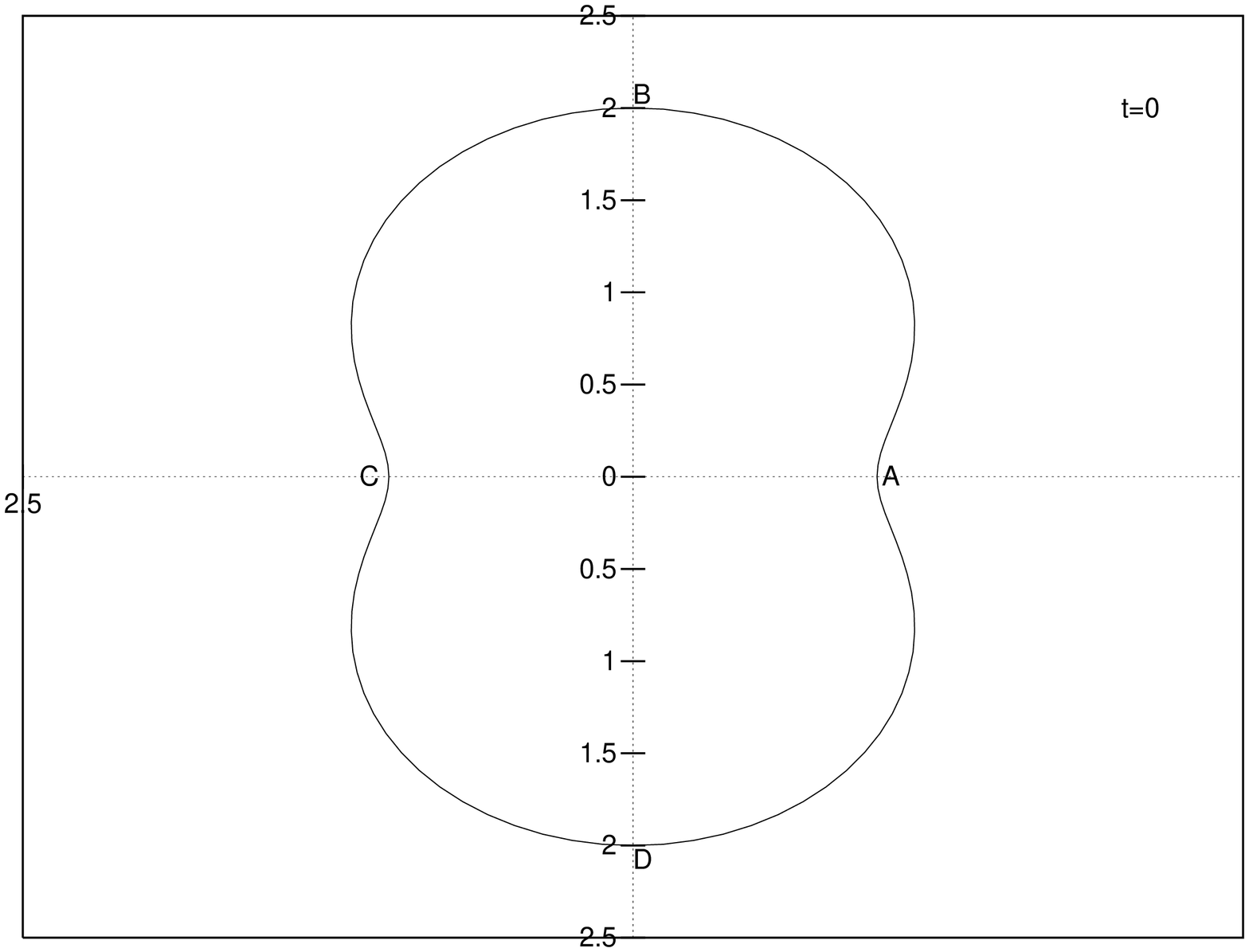,height=8cm,width=6cm,angle=-90}  \\
\large \bf (b) \rm \normalsize			&	
\epsfig{file=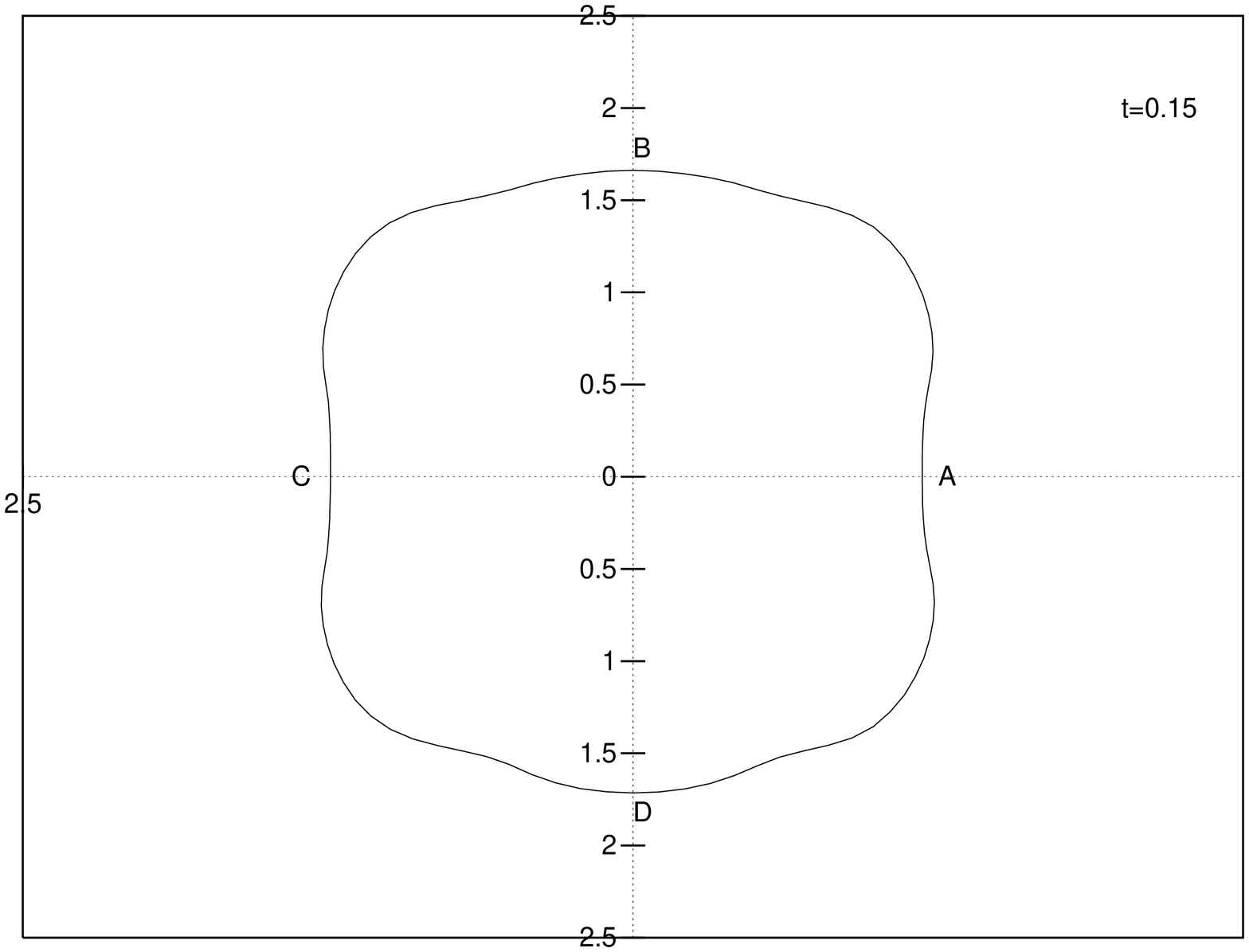,height=8cm,width=6cm,angle=-90}  \\
\large \bf (c) \rm \normalsize			&	
\epsfig{file=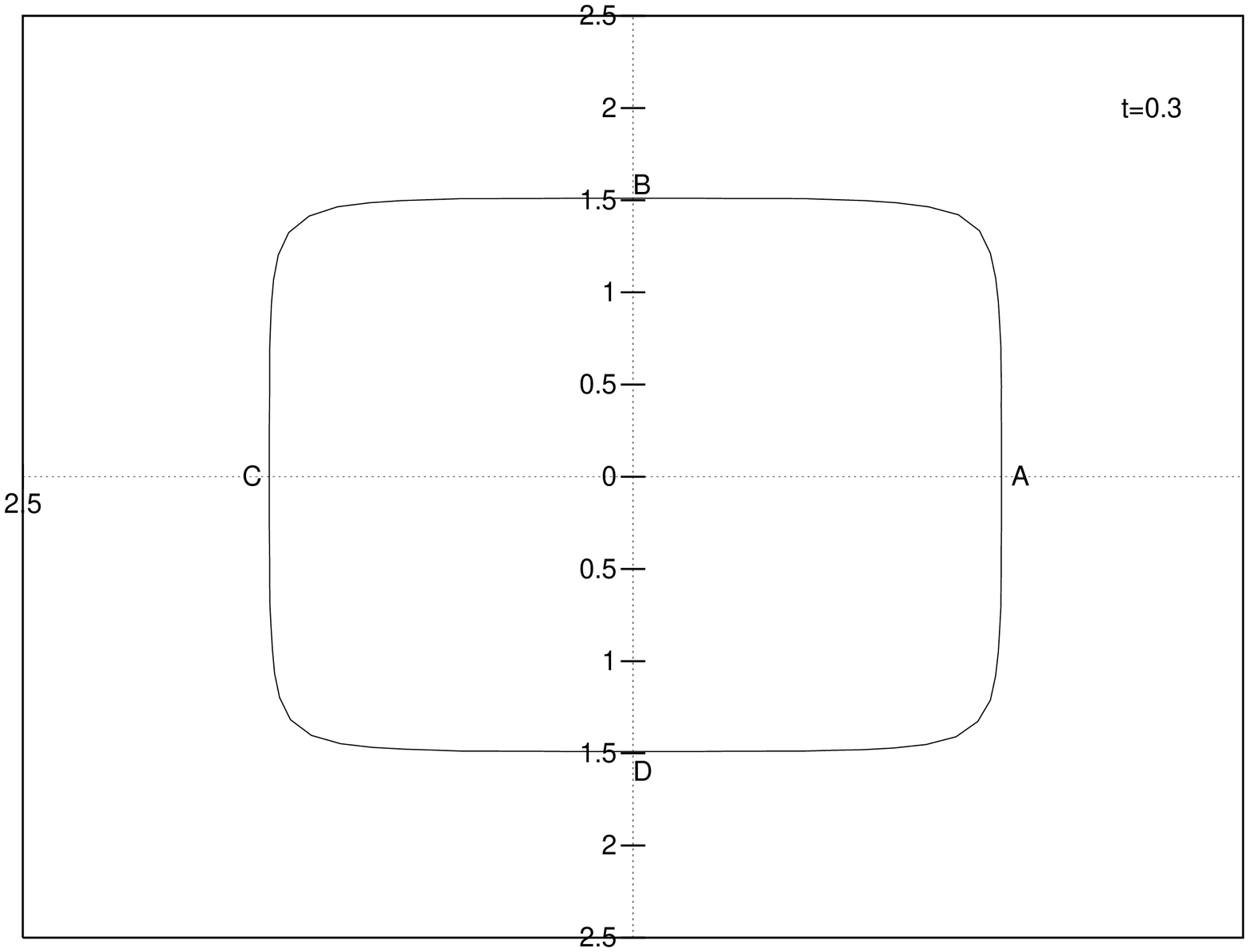,height=8cm,width=6cm,angle=-90}  
\end{tabular}
\caption{}
\label{front-06}
\end{figure}

In this manuscript we have derived equation (\ref{eq:flwmcextc}) which
governs the evolution of a fully developed front in a reaction-diffusion
system described by (\ref{eq:def1f}) when \( \epsl \ll 1 \) (slow diffusion, 
large separation between sites or fast reaction). This equation generalizes 
the FMC equation (\ref{eq:flwmc}) to include the effects of stronger 
nonlinearities and accounts for the influence of the non-homogeneous reaction
term on the motion of the interface. The motion of fronts according to 
(\ref{eq:flwmcextc}) is qualitatively different from that of the homogeneous 
nonlinear reaction term counterpart given by the FMC equation, a phenomenon 
that was pointed out by Keener \cite{kn:kee2}. This difference arises primarily
from the fact that the function \( \beta \) acts as a "potential" function for
the motion of the front. For the one dimensional case, an initial front 
initially placed between two maxima of \( \beta \) (which for a homogeneous 
nonlinear reaction term will move with a velocity proportional to \( \bnu \)) 
asymptotically approaches the intervening minimum. For the radially symmetric 
two-dimensional case circular domains of one phase inside the other are 
stabilized. We found numerically that other closed curves present the same 
phenomenon. These results (Fig. \ref{front-04}-\ref{front-06}) suggest that
the curvature of the front may play a role in "balancing" the "force" exerted 
by the function \( \beta \) on the front. Analytical and numerical work will 
be necessary in order to elucidate this effect, which may be crucial in the 
selection of the equilibrium pattern.
\bk  

\nd In equation (\ref{eq:def1}) or (\ref{eq:def1f}) the function \( \beta \)
can be choosen to depend not only on the spatial variable but also on 
\( t \). The fire-diffuse-fire model of dynamics of intracellular calcium 
waves \cite{kn:ponkei1,kn:peapon1} is of this type. In this case equation
(\ref{eq:flwmcextc}) will still govern the evolution of fronts, where now
\( \beta = \beta(x,y,t) \). The form of \( \beta(x,y,t) \) will depend, of 
course, on the particular model. One might, for example, have the product of a
spatially dependent function \( \beta(x,y) \) with a probabilistic time 
dependent function.
\bk

\nd The results presented here have implications for the selection of patterns
in systems of the type

\begeq
	\left\{ \begin{array}{l}	
		\eps2\ \phit = \eps2\ \lap-phi + u\ [ f(\phi) + \epsl\ h ], \\
									    \\
		\tau\ u_{t} = D_{u}\ \Delta u + g(u,\phi), 
							\label{eq:system}
		\end{array}
	\right.
\endeq

\nd where \( D_{u} \) is a diffusion constant, \( g \) is a given nonlinear
function the time constant, \( \tau \), is assumed to be large. If \( u \)
rapidly approaches a  non-homogeneous steady state, then depending on the 
initial conditions, it may induce a non-homogeneous steady state in \( \phi \) 
where \( u \) acts as the function \( \beta \) in (\ref{eq:def1f}). 
\bk
\bk
\bk

\nd We thank Boris Malomed and Igor Mitkov for reading the manuscript, 
constructive criticism and useful comments.

\bibliographystyle{unsrt}
\bibliography{tesis}

\newpage

\nd \underline{Captions}
\bk  

\begin{figure}[ph]
\begin{tabular}{llllllll}
\large \bf (a) \rm \normalsize			&
\epsfig{file=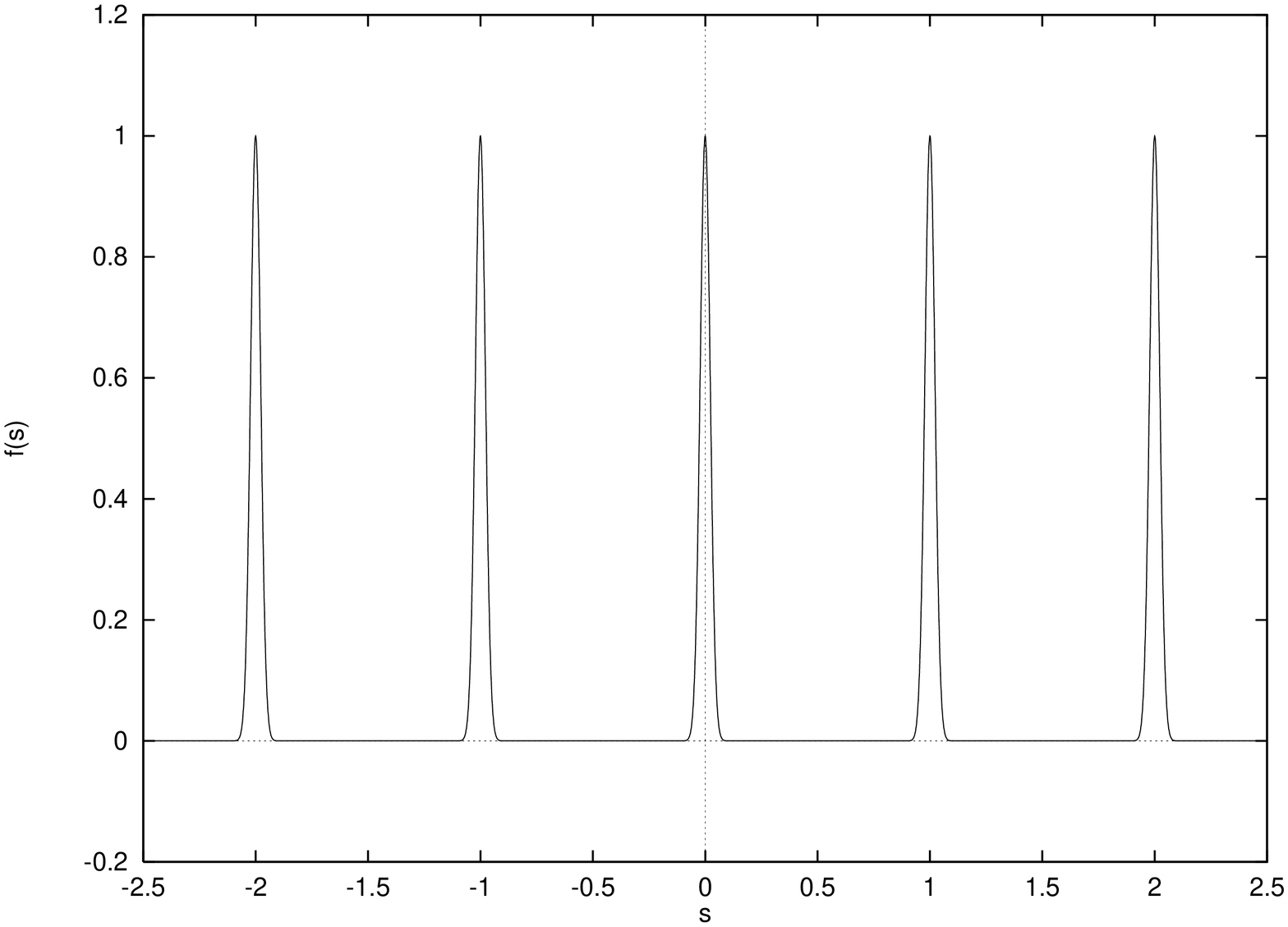,height=10cm,width=4.5cm,angle=-90}  \\
\large \bf (b) \rm \normalsize			&	
\epsfig{file=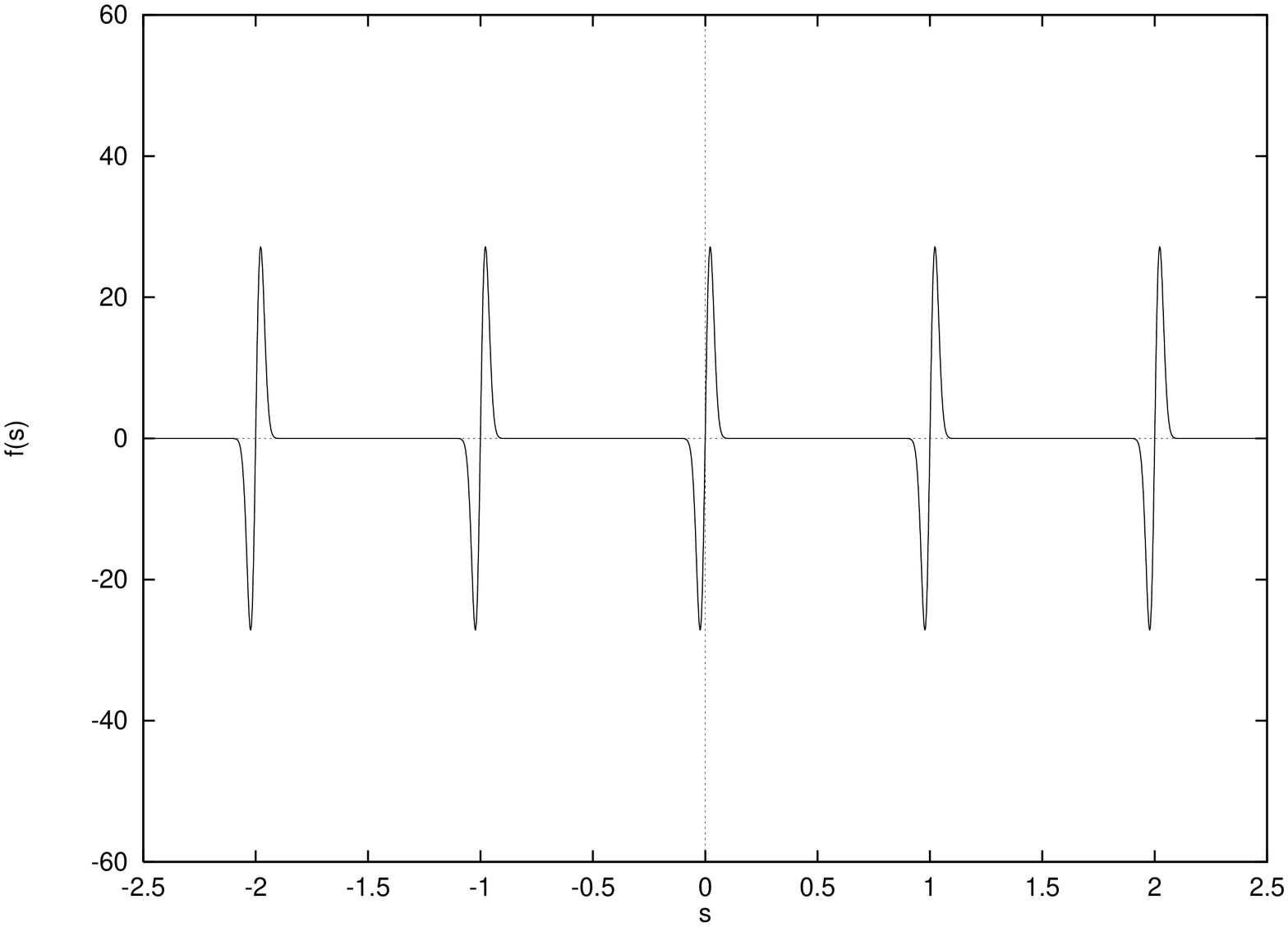,height=10cm,width=4.5cm,angle=-90}  \\
\large \bf (c) \rm \normalsize			&	
\epsfig{file=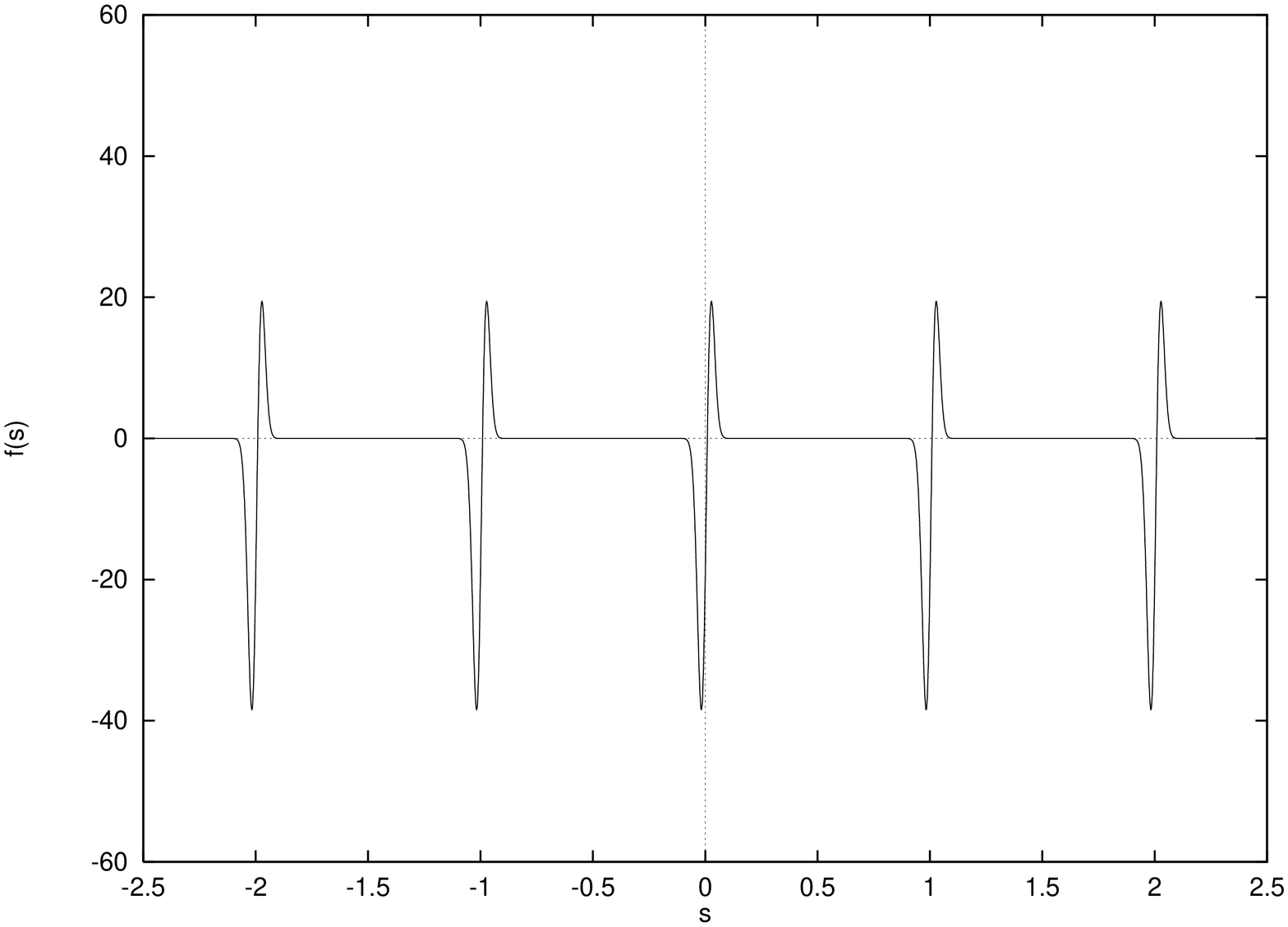,height=10cm,width=4.5cm,angle=-90}  \\
\large \bf (d) \rm \normalsize			&	
\epsfig{file=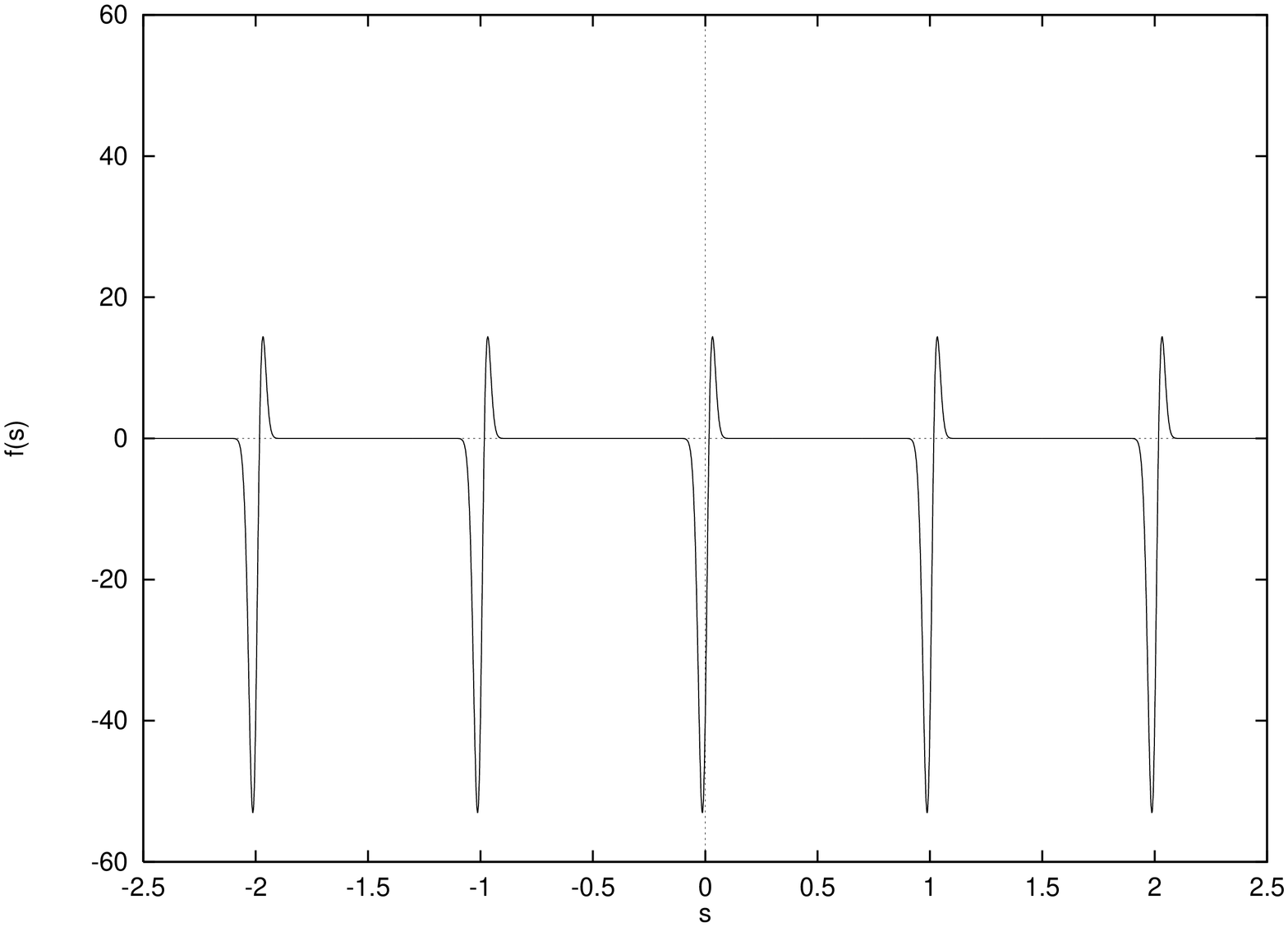,height=10cm,width=4.5cm,angle=-90}  
\end{tabular}
\caption{}
\label{discsites-1}
\end{figure}

\nd Figure \ref{discsites-1}: 

\nd a) Graph of \( \beta(s) = \sum_{k=1}^{5}\ e^{-1000 (s-x_{k})^{2}}
\) for \( x_{1} = 2 \), \( x_{2} = 1 \), \( x_{3} = 0 \) and 
\( x_{4} = -1 \), \( x_{5} = -2 \).

\nd b) Graph of \( g(s) = - \bet'(s) - 2\ \beta^{\frac{3}{2}}(s)\ \bar{h}
\) for \( \bet(s) \) as in (a) and \( h = 0 \).

\nd c) Graph of \( g(s) = - \bet'(s) - 2\ \beta^{\frac{3}{2}}(s)\ \bar{h}
\) for \( \bet(s) \) as in (a) and \( h = 10 \).

\nd d) Graph of \( g(s) = - \bet'(s) - 2\ \beta^{\frac{3}{2}}(s)\ \bar{h}
\) for \( \bet(s) \) as in (a) and \( h = 20 \).

\begin{figure}[ph]
\begin{tabular}{llllllll}
\large \bf (a) \rm \normalsize			&
\epsfig{file=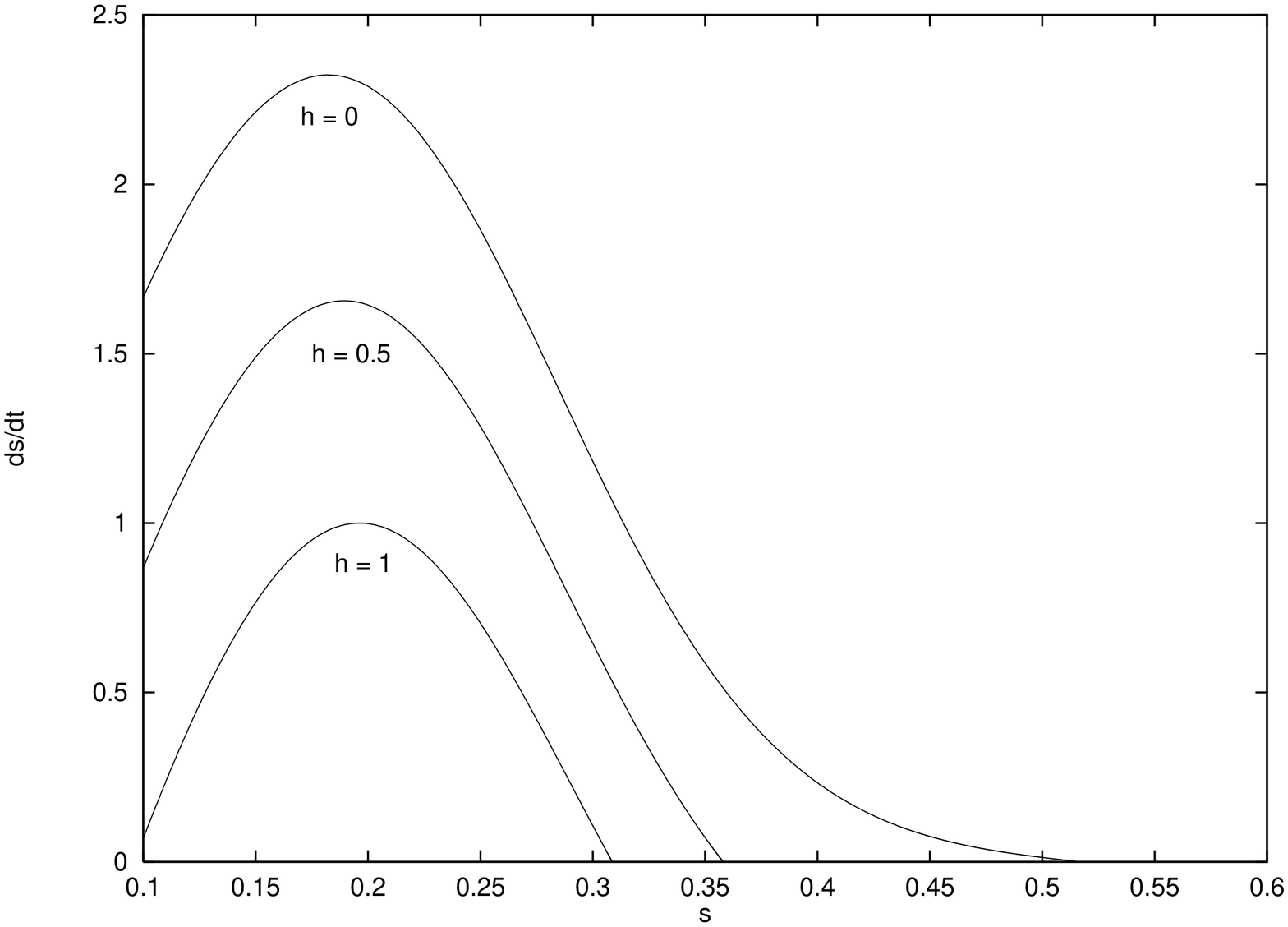,height=12cm,width=6cm,angle=-90}  \\
\large \bf (b) \rm \normalsize			&	
\epsfig{file=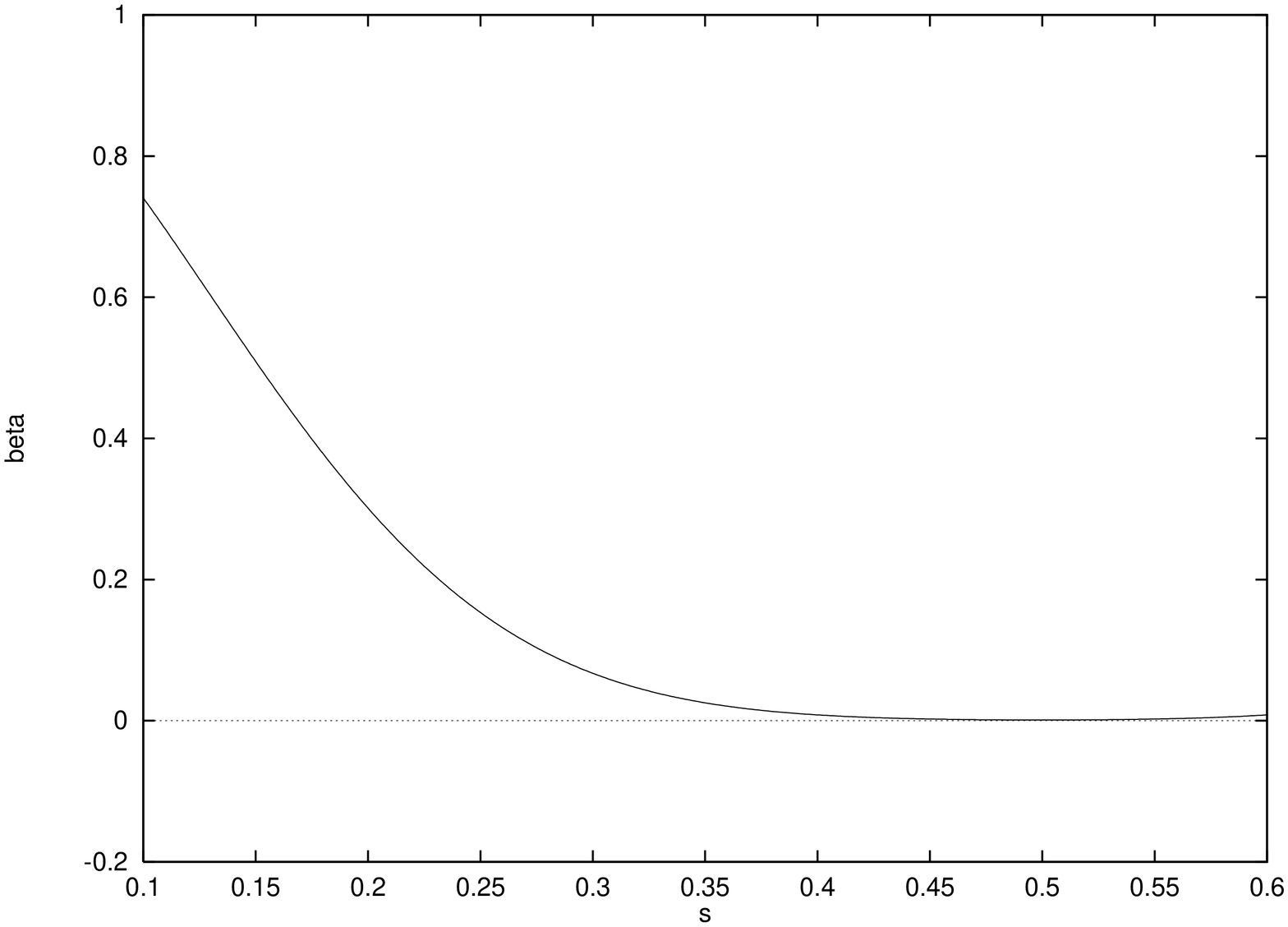,height=12cm,width=6cm,angle=-90}  \\
\large \bf (c) \rm \normalsize			&	
\epsfig{file=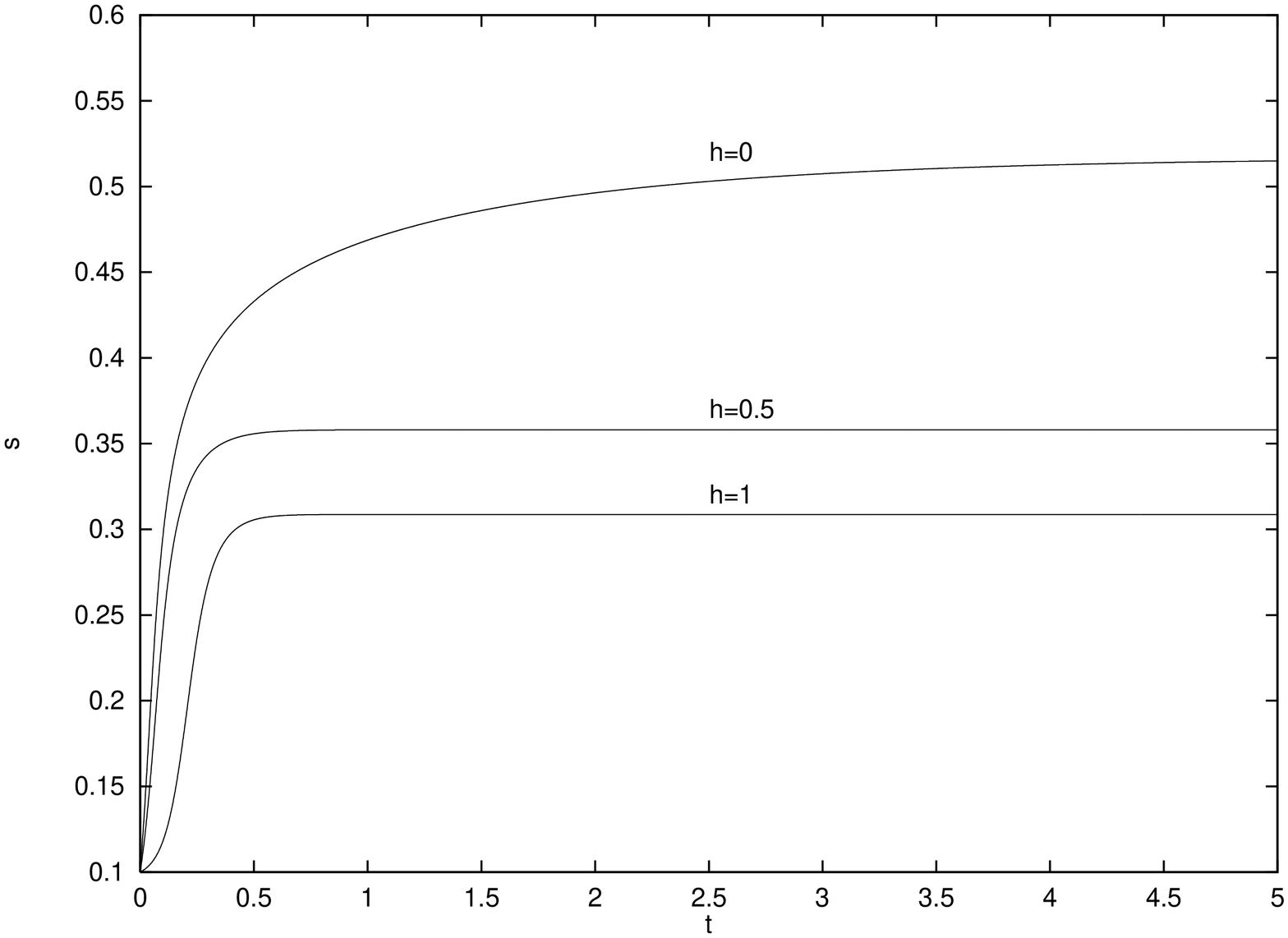,height=12cm,width=6cm,angle=-90}  
\end{tabular}
\caption{}
\label{front-1}
\end{figure}

\nd Figure \ref{front-1}: 

\nd a) Dependence of the front velocity \( s_{t} \) on the position of
the front \( s \)  in equation (\ref{eq:flow1d})

\nd b) Dependence of \( \bet(s) \) on the position of the front \( s \).

\nd c) Dependence of the position of the front \( s \) on time \( t \). 

\nd In all the cases \( \bet(s) \) given by 
(\ref{eq:bet1d}) with \( N = 2 \), \( \eta = 30 \), \( \Delta t =
0.0001 \) and \( \bnu = 0, 0.5, 1 \) from above,. The front is initially
in \( s_{0} = 0.1 \).
\bk

\begin{figure}[ph]
\begin{tabular}{llllllll}
\large \bf (a) \rm \normalsize			&
\epsfig{file=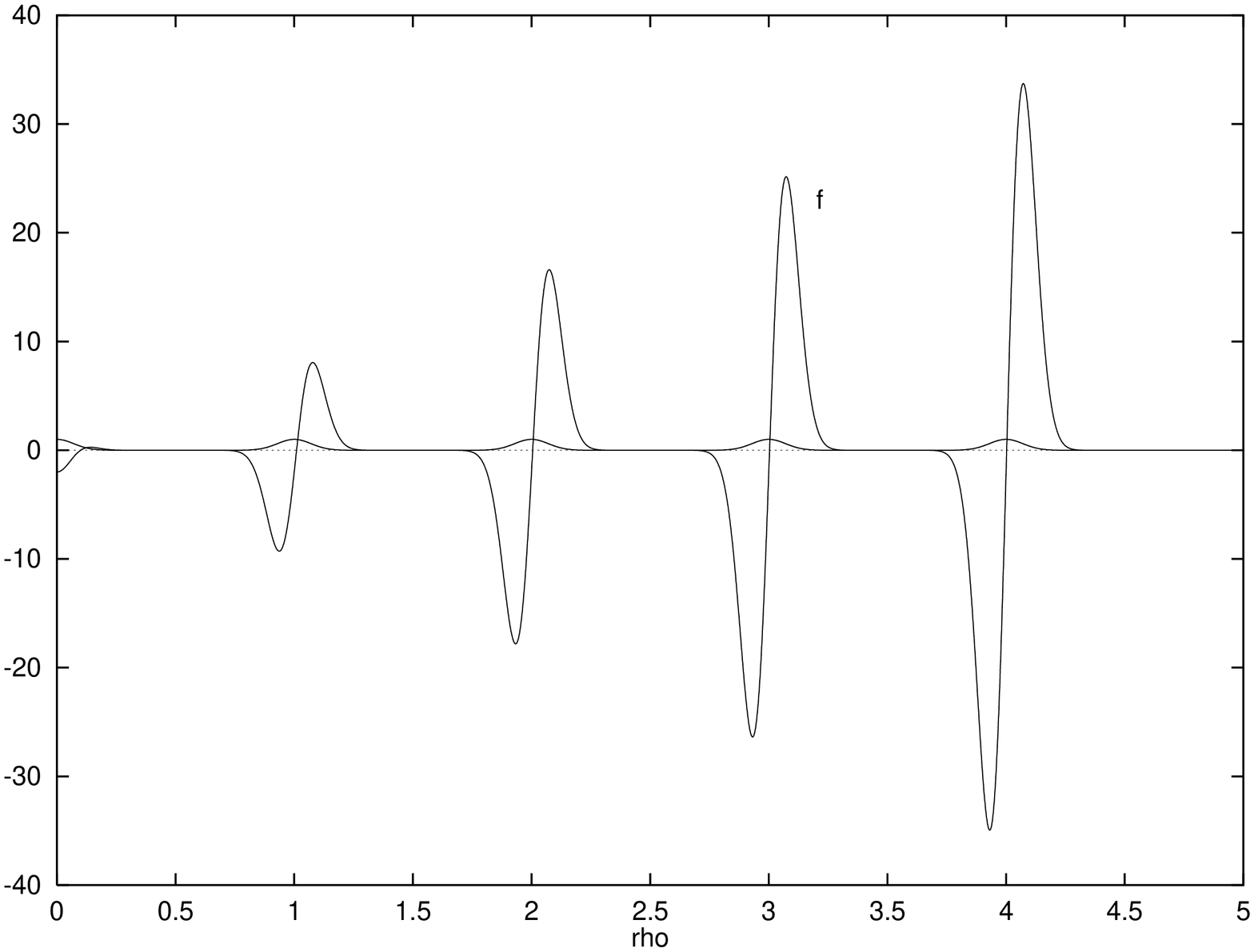,height=7cm,width=8cm,angle=-90}  &
\large \bf (b) \rm \normalsize			&	
\epsfig{file=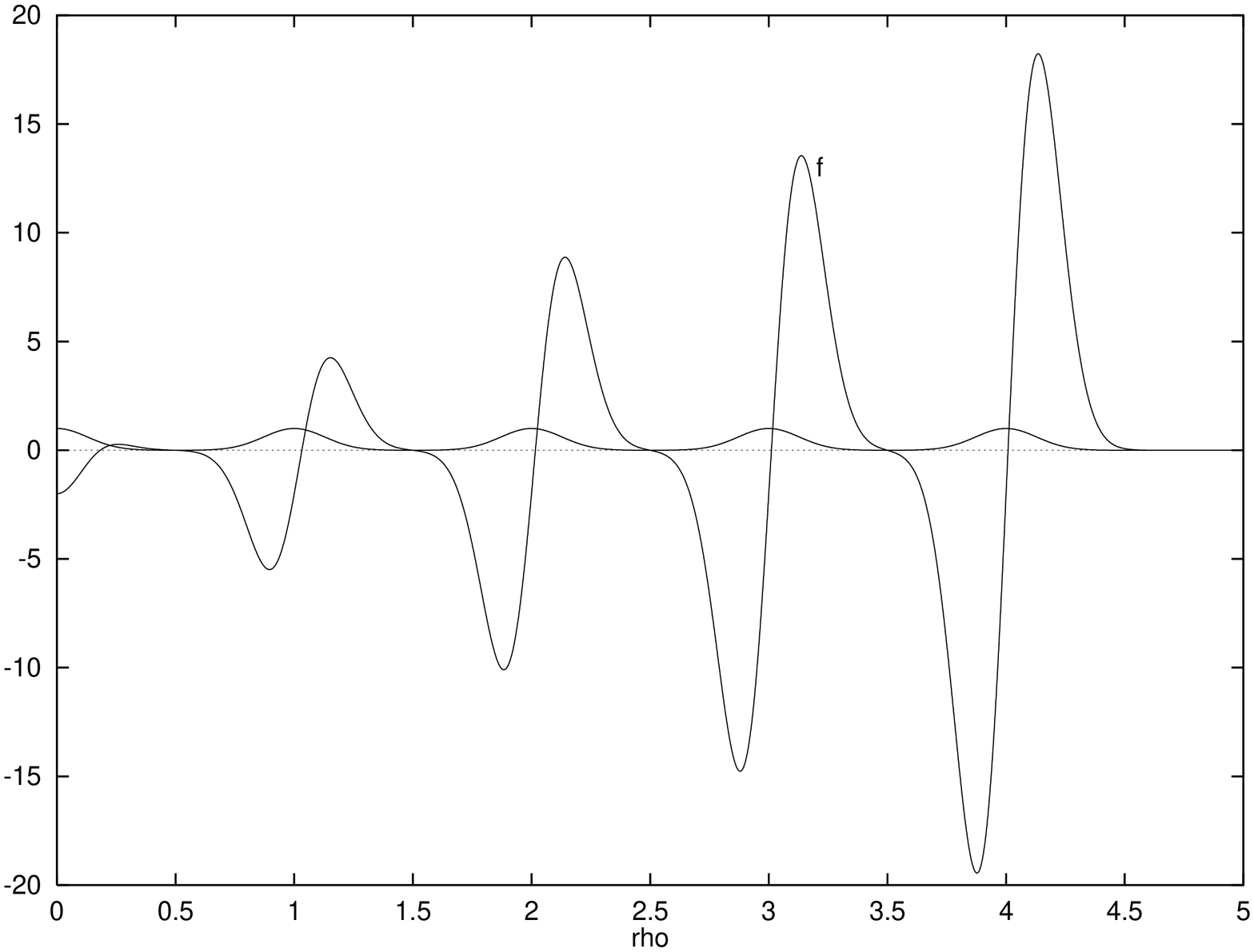,height=7cm,width=8cm,angle=-90}  \\
\large \bf (c) \rm \normalsize			&	
\epsfig{file=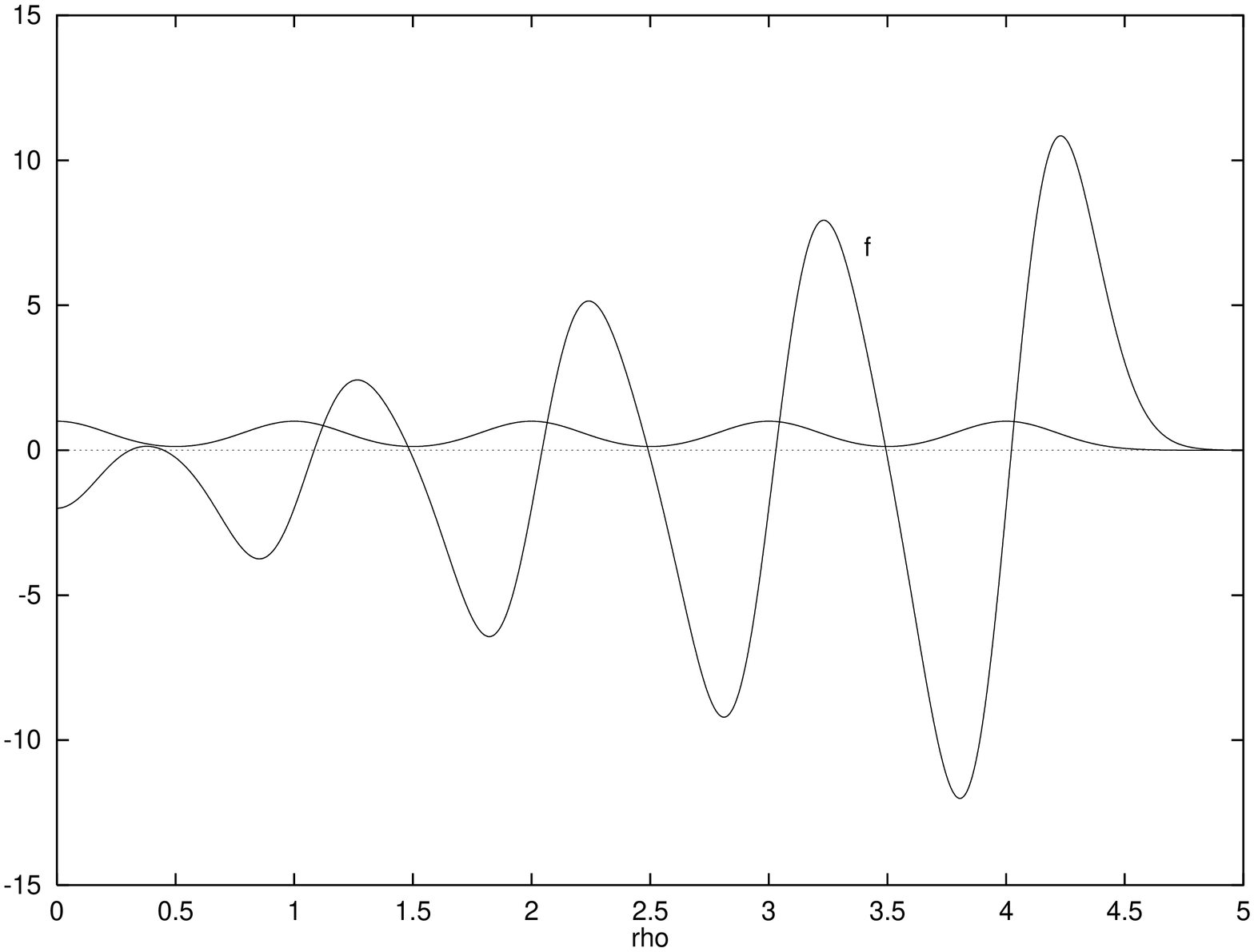,height=7cm,width=8cm,angle=-90}  &
\large \bf (d) \rm \normalsize			&	
\epsfig{file=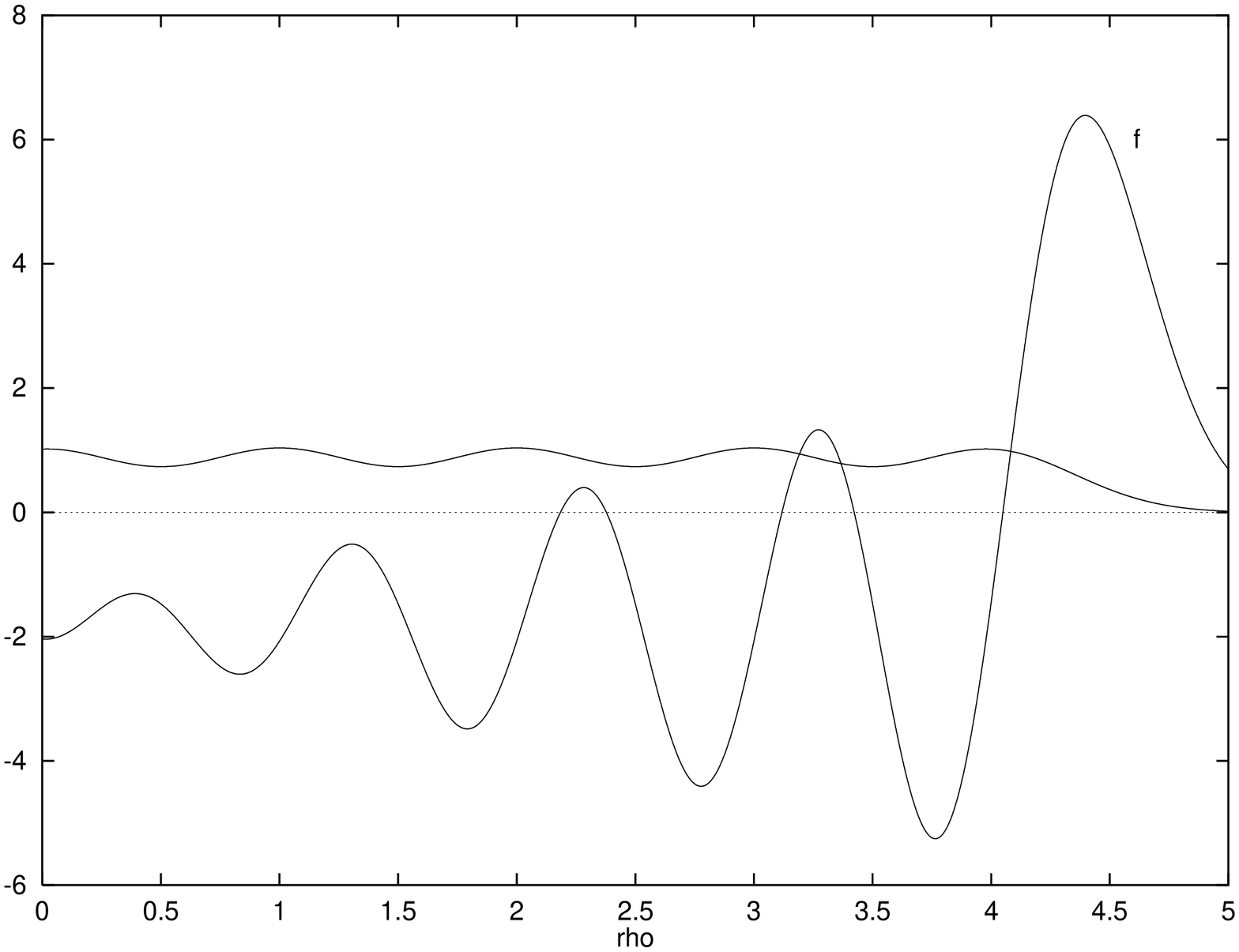,height=7cm,width=8cm,angle=-90}  
\end{tabular}
\caption{}
\label{discsites-2}
\end{figure}

\nd Figure \ref{discsites-2}: 

\nd Dependence of \( \bet(\rho) = \sum_{k=1}^{5}\ e^{-\eta 
(\rho-\rho_{k})^{2}} \) and \( f(\rho) - 2\ \bet(\rho) - 
\rho\ \bet_{\rho}(\rho) \) for \( \rho_{1} = 0 \), \( \rho_{2} = 1 \), 
\( \rho_{3} = 2 \), \( \rho_{4} = 3 \), \( \rho_{5} = 4 \) and 

\nd a) \( \eta = 100 \).

\nd b) \( \eta = 30 \).

\nd c) \( \eta = 11 \).

\nd d) \( \eta = 4 \).
\bk

\begin{figure}[ph]
\begin{tabular}{llllllll}
\large \bf (a) \rm \normalsize			&
\epsfig{file=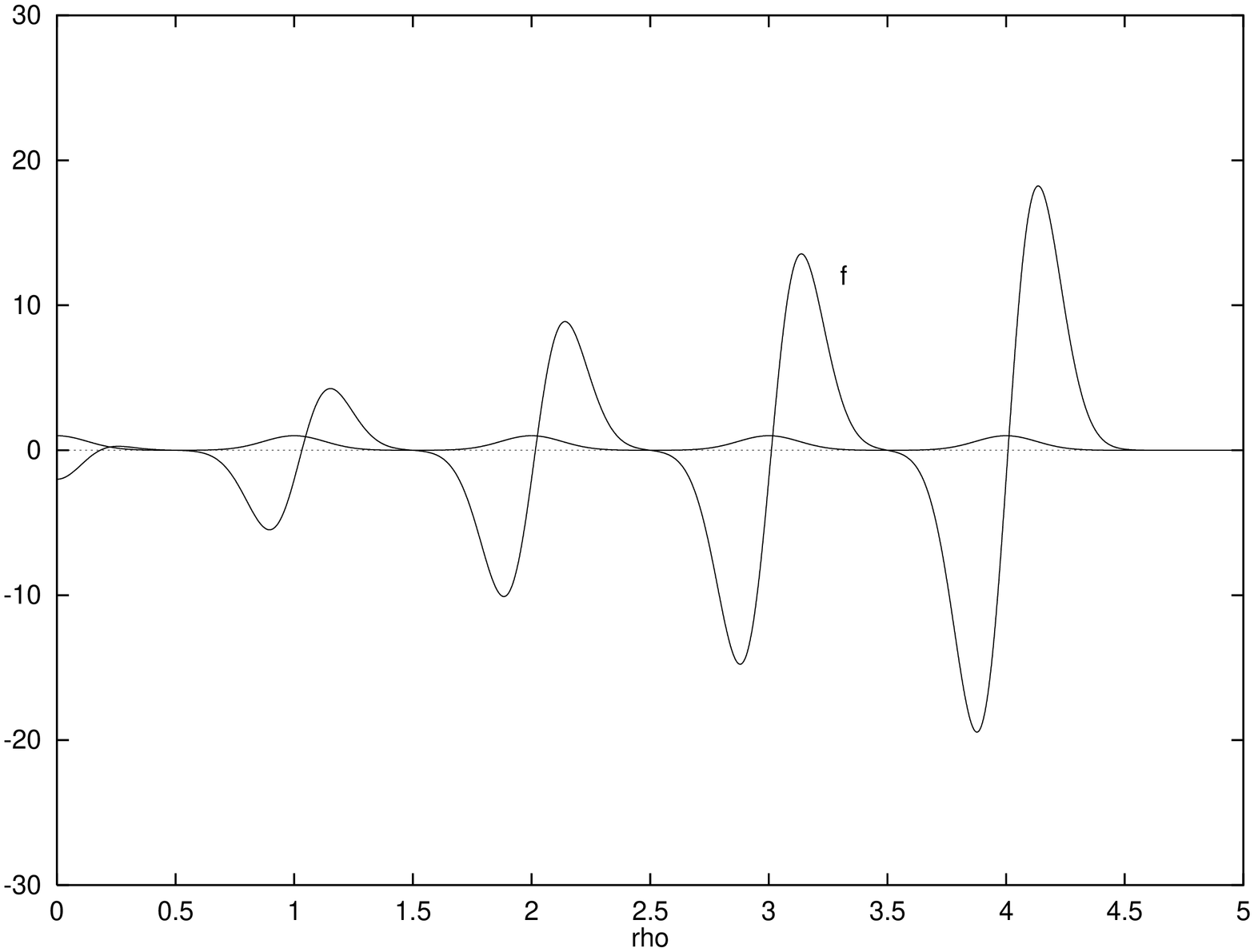,height=7cm,width=8cm,angle=-90}  &
\large \bf (b) \rm \normalsize			&	
\epsfig{file=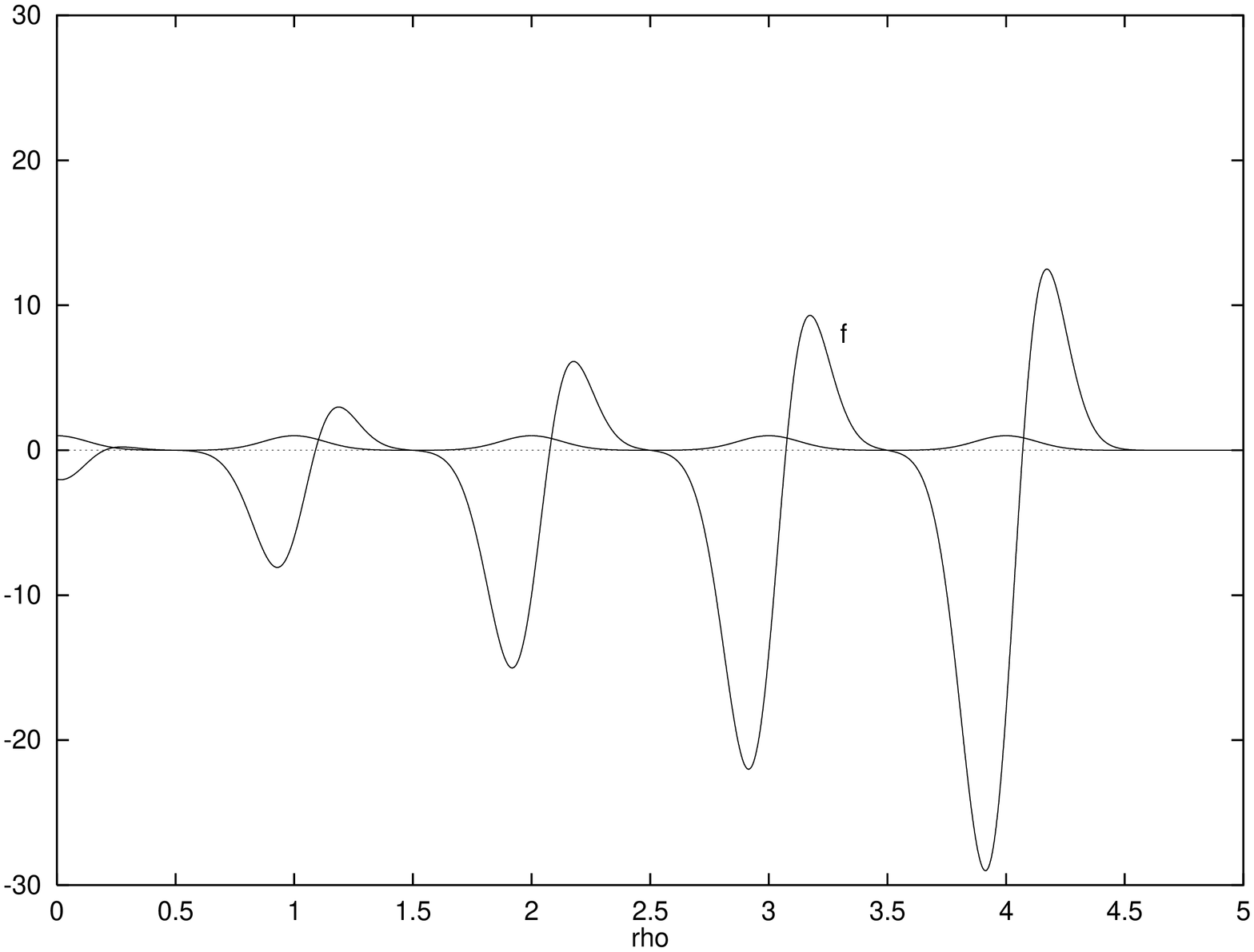,height=7cm,width=8cm,angle=-90}  \\
\large \bf (c) \rm \normalsize			&	
\epsfig{file=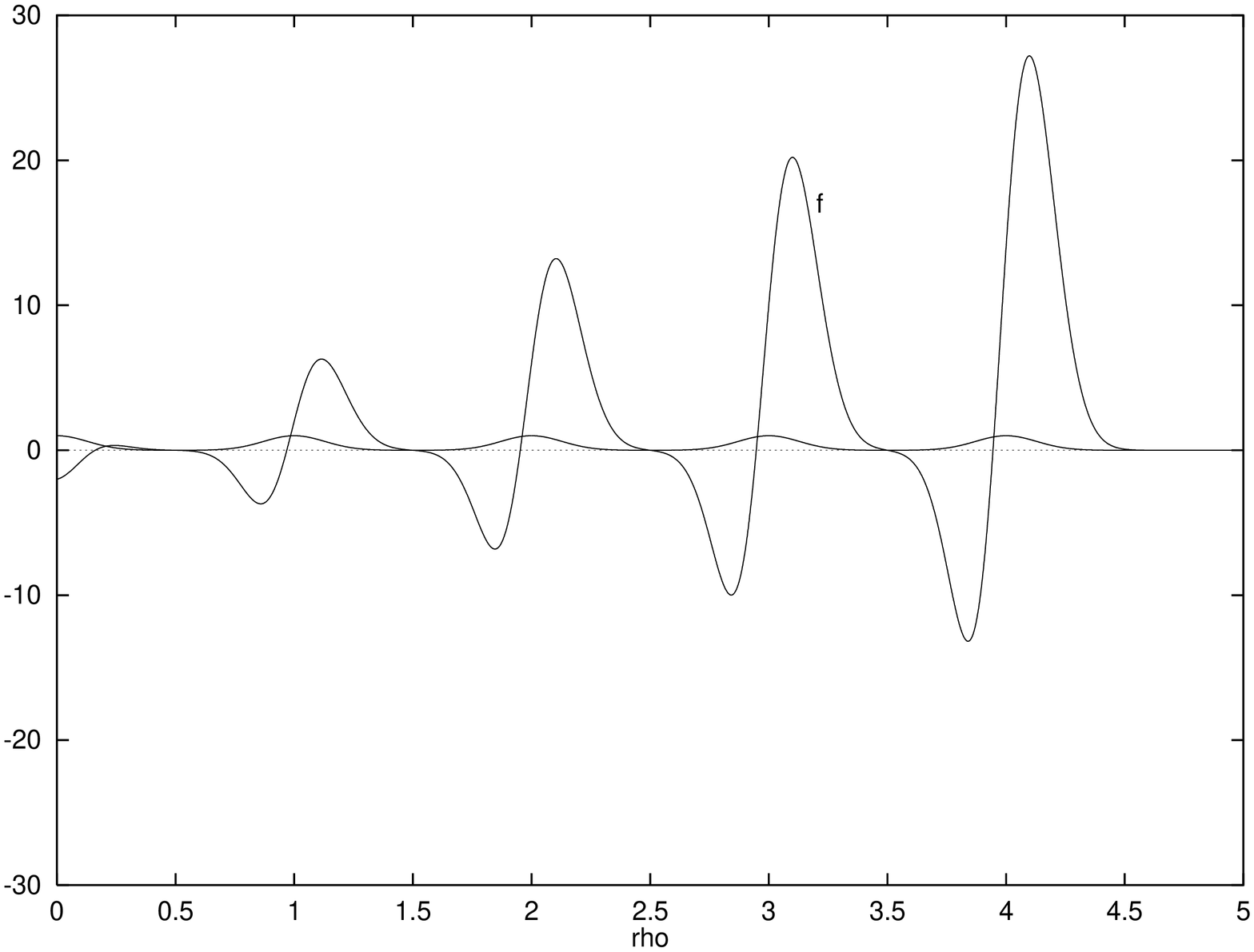,height=7cm,width=8cm,angle=-90}  
\end{tabular}
\caption{}
\label{discsites-3}
\end{figure}

\nd Figure \ref{discsites-3}: 

\nd Dependence of \( \bet(\rho) = \sum_{k=1}^{5}\ e^{-\eta 
(\rho-\rho_{k})^{2}} \) and \( f(\rho) - 2\ \bet(\rho) - 
\rho\ \bet_{\rho}(\rho) - 2\ \rho\ \bet^{\frac{3}{2}}(\rho)\ \bnu \) 
for \( \rho_{1} = 0 \), \( \rho_{2} = 1 \), \( \rho_{3} = 2 \), 
\( \rho_{4} = 3 \), \( \rho_{5} = 4 \), \( \eta = 30 \) and 

\nd a) \( \bnu = 0 \)

\nd b) \( \bnu = 2 \)

\nd c) \( \bnu = -2 \)
\bk

\begin{figure}[ph]
\begin{tabular}{llllllll}
\epsfig{file=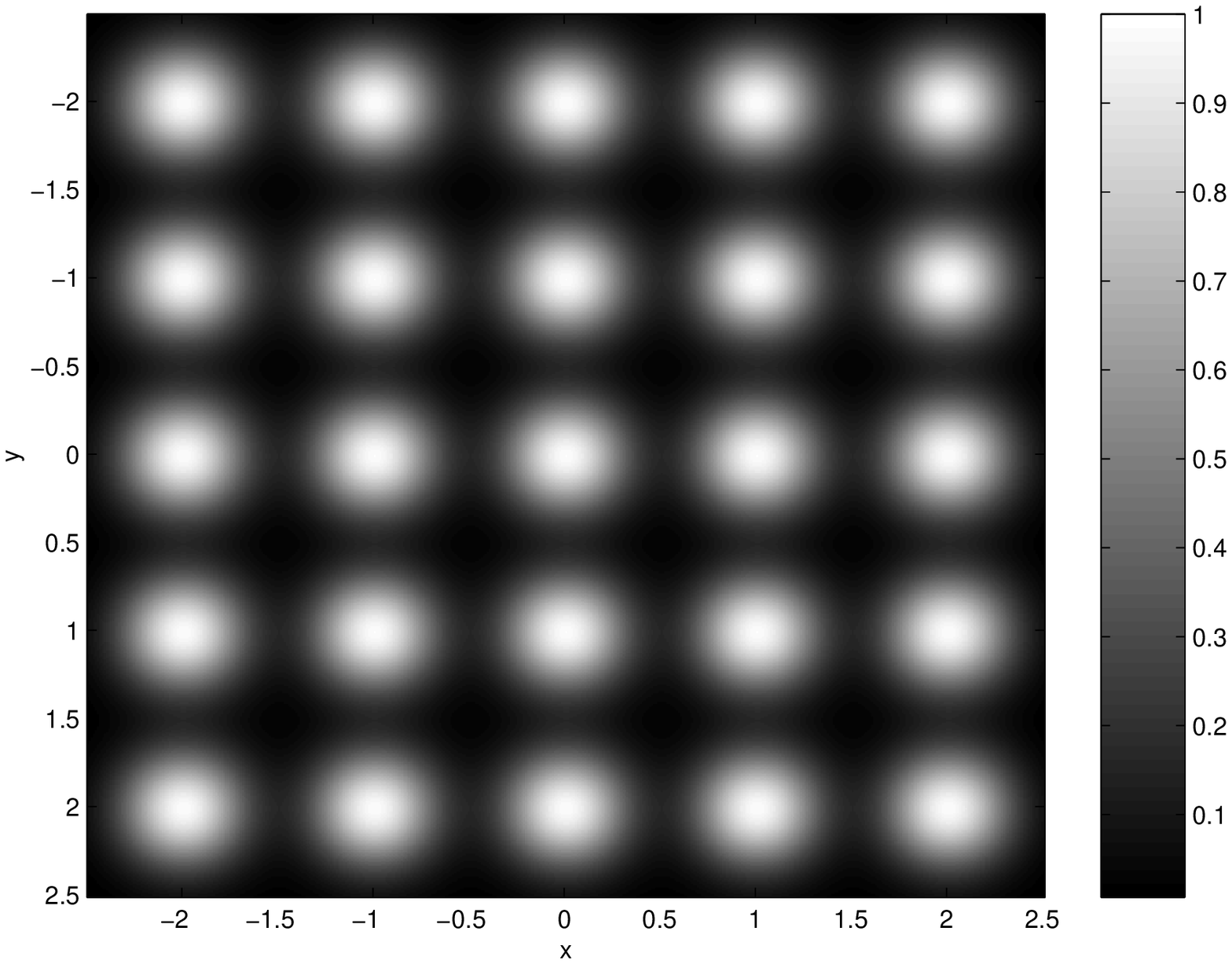,height=12cm,width=15cm,angle=0}  
\end{tabular}
\caption{}
\label{front-2d-04}
\end{figure}

\nd Figure \ref{front-2d-04}

\nd Graph of \( \bet(x,y) = \sum_{k=1}^{5}\ \sum_{j=1}^{5}
\sigma(x-x_{k},y-y_{j}) \) where \( \sigma(x,y) = e^{-10\ (x^{2}+y^{2})} \)
with \( (x_{k},y_{j}) \), \( k, j = -2, \ldots, 2 \) and \(
(x_{0},y_{0}) = (0,0) \).   
\bk

\nd Figure \ref{front-04}

\nd Graphs of the evolution of a circle with initial radius equal to 2 
according to (\ref{eq:flwmcextc}) with 
\( \bnu = 0 \) and \( \bet(x,y) = \sum_{k=1}^{5}\ \sum_{j=1}^{5}
\sigma(x-x_{k},y-y_{j}) \) where \( \sigma(x,y) = e^{-10\ (x^{2}+y^{2})} \)
with \( (x_{k},y_{j}) \), \( k, j = -2, \ldots, 2 \) and \(
(x_{0},y_{0}) = (0,0) \), for

\nd a) \( t = 0 \)

\nd b) \( t = 0.15 \)

\nd c) \( t = 0.3 \)
\bk

\nd Figure \ref{front-05}

\nd Graphs of the evolution of an ellipse, \( \frac{x^{2}}{4} + y^{2} = 1 \) 
according to (\ref{eq:flwmcextc}) with \( \bnu = 0 \) and 
\( \bet(x,y) = \sum_{k=1}^{5}\ \sum_{j=1}^{5}
\sigma(x-x_{k},y-y_{j}) \) where \( \sigma(x,y) = e^{-10\ (x^{2}+y^{2})} \)
with \( (x_{k},y_{j}) \), \( k, j = -2, \ldots, 2 \) and \(
(x_{0},y_{0}) = (0,0) \), for

\nd a) \( t = 0 \)

\nd b) \( t = 0.15 \)

\nd c) \( t = 0.4 \)
\bk

\nd Figure \ref{front-06}

\nd Graphs of the evolution of the function \( (cos(\tet),2 sin(\tet)) \)
according to (\ref{eq:flwmcextc}) with \( \bnu = 0 \) 
\( \bet(x,y) = \sum_{k=1}^{5}\ \sum_{j=1}^{5}
\sigma(x-x_{k},y-y_{j}) \) where \( \sigma(x,y) = e^{-10\ (x^{2}+y^{2})} \)
with \( (x_{k},y_{j}) \), \( k, j = -2, \ldots, 2 \) and \(
(x_{0},y_{0}) = (0,0) \), for

\nd a) \( t = 0 \)

\nd b) \( t = 0.15 \)

\nd c) \( t = 0.6 \)
\bk

\end{document}